\documentclass[12pt,preprint]{aastex}

\newcommand   {\mic}   {\mbox{$\mu$m}}

\renewcommand {\deg}   {\mbox{$^\circ$}}

\newcommand   {\arcs}  {\mbox{$^{\prime\prime}$}}

\newcommand   {\kms}   {\mbox{km\,s$^{-1}$}}
\renewcommand {\ga}    {\mbox{\rlap{\hbox{\lower5pt\hbox{$\sim$}}}\hbox{$>$}}}
\renewcommand {\la}    {\mbox{\rlap{\hbox{\lower5pt\hbox{$\sim$}}}\hbox{$<$}}}

\begin{document}



\def\kms {\hbox{km{\hskip0.1em}s$^{-1}$}} 
\def\msol{\hbox{$\hbox{M}_\odot$}}
\def\lsol{\hbox{$\hbox{L}_\odot$}}
\def\kms{km s$^{-1}$}
\def\Blos{B$_{\rm los}$}
\def\etal   {{\it et al. }}                     
\def\psec           {$.\negthinspace^{s}$}
\def\pasec          {$.\negthinspace^{\prime\prime}$}
\def\pdeg           {$.\kern-.25em ^{^\circ}$}
\def\degree{\ifmmode{^\circ} \else{$^\circ$}\fi}
\def\ee #1 {\times 10^{#1}}          
\def\ut #1 #2 { \, \textrm{#1}^{#2}} 
\def\u #1 { \, \textrm{#1}}          
\def\nH {n_\mathrm{H}}

\def\ddeg   {\hbox{$.\!\!^\circ$}}              
\def\deg    {$^{\circ}$}                        
\def\le     {$\leq$}                            
\def\sec    {$^{\rm s}$}                        
\def\msol   {\hbox{$M_\odot$}}                  
\def\i      {\hbox{\it I}}                      
\def\v      {\hbox{\it V}}                      
\def\dasec  {\hbox{$.\!\!^{\prime\prime}$}}     
\def\asec   {$^{\prime\prime}$}                 
\def\dasec  {\hbox{$.\!\!^{\prime\prime}$}}     
\def\dsec   {\hbox{$.\!\!^{\rm s}$}}            
\def\min    {$^{\rm m}$}                        
\def\hour   {$^{\rm h}$}                        
\def\amin   {$^{\prime}$}                       
\def\lsol{\, \hbox{$\hbox{L}_\odot$}}
\def\sec    {$^{\rm s}$}                        
\def\etal   {{\it et al. }}                     
\def\Chandra   {{\it Chandra }}
\def\xbar   {\hbox{$\overline{\rm x}$}}         

\slugcomment{ Publication date, July  20. 2008}
\shorttitle{}
\shortauthors{}

\title{Simultaneous Chandra, CSO
and VLA Observations of Sgr A*: The Nature of  Flaring Activity}

\author{F. Yusef-Zadeh\altaffilmark{1},
M. Wardle\altaffilmark{2},
C. Heinke\altaffilmark{3},
C. D. Dowell\altaffilmark{4},
D. Roberts\altaffilmark{1}, 
F. K. Baganoff\altaffilmark{5} \&
W. D. Cotton\altaffilmark{6}
}
\altaffiltext{1}{Department of Physics and Astronomy,
Northwestern University, Evanston, Il. 60208
(zadeh@northwestern.edu)}
\altaffiltext{2}{Department of Physics, Macquarie University, Sydney NSW 2109,
Australia (wardle@physics.mq.edu.au)}
\altaffiltext{3}{Department of Astronomy, 
University of Virginia (coh5z@virginia.edu)}
\altaffiltext{4}{ Cal Tech, Jet Propulsion Laboratory, Pasadena, CA 91109 
(cdd@submm.caltech.edu)}
\altaffiltext{5}{Kavli Institute for Astrophysics and Space Research, MIT,
Cambridge,MA 02139-4307  (fkb@space.mit.edu)}
\altaffiltext{6}{National Radio Astronomy Observatory, 520 
Edgemont Road, Charlottesville, VA 22903-2475\ (bcotton@nrao.edu)}


\begin{abstract} 
Sgr A*, the massive black hole at the center of the Galaxy, varies in
radio through X-ray emission on hourly time scales.  The flare
activity is thought to arise from the innermost region of an accretion
flow onto Sgr A*.  We present simultaneous light curves of Sgr A* in
radio, sub-mm and X-rays that show a possible time delay of 110$\pm17$
minutes between X-ray and 850 $\mu$m suggesting that the sub-mm flare
emission is optically thick.  At radio wavelengths, we detect time
lags of of $20.4\pm6.8, 30\pm12$ and 20$\pm6$ minutes between the
flare peaks observed at 13 and 7 mm (22 and 43 GHz) in three different
epochs using the VLA. Linear polarization of 1$\pm0.2$\% and
0.7$\pm0.1$\% is detected at 7 and 13 mm, respectively, when averaged
over the entire observation on 2006 July 17.  A simple model of a
bubble of synchrotron emitting electrons cooling via adiabatic
expansion can explain the time delay between various wavelengths, the
asymmetric shape of the light curves, and the observed polarization of
the flare emission at 43 and 22 GHz.  The derived physical quantities
that characterize the emission give an expansion speed of v$_{exp}
\sim 0.003-0.1$c, magnetic field of B$\sim$10-70 Gauss and particle
spectral index p$\sim$1-2.  These parameters suggest that the
associated plasma cannot escape from Sgr A* unless it has a large bulk
motion.
\end{abstract}

\keywords{accretion, accretion disks --- black hole physics ---
 Galaxy: center}

\section{Introduction}
\label{introduction} 

Recent observations provide compelling evidence that the compact
nonthermal radio source Sgr A* can be identified with a massive black
hole at the center of the Galaxy.  A major breakthrough in our
understanding of this source came from stellar orbital
measurements showing a mass 3-4 $\times 10^6$ \msol\ within 45 AU of
the position of Sgr A* (e.g., Sch\"odel et al.  2003; Ghez et al.
2004; Eisenhauer et al.  2005).  This dark, massive object has been
uniquely identified with the radio source Sgr A* through limits on the
proper motion of the radio source, which show that Sgr A* must contain
$>4\times10^{5}$ \msol\ and does not orbit another massive object
(Reid \& Brunthaler 2004).

A ``sub-mm bump'' was detected in the broad-band spectrum of Sgr A*
(Zylka et al.  1995; Serabyn et al.  1997; Falcke et al.  1998); the
sub-mm peak is thought to reflect the transition between optically
thin emission at higher frequencies, and optically thick emission at
lower frequencies.  Observations of the time variability and
polarization (Bower et al.  2005; Marrone et al.  2006; Macquart \&
Bower 2006; Yusef-Zadeh et al.  2007) are providing
additional opportunities to study the innermost regions, within just a
few Schwarzschild radii of the black hole event horizon, a region
currently inaccessible via high spatial resolution imaging.

The energy radiated by Sgr A* is thought to be liberated from gas
falling towards the black hole after being captured from the powerful
winds of members of its neighboring cluster of massive stars (e.g.,
Melia 1992).  The radiated power is several orders of magnitudes lower
than expected, prompting a number of theoretical models to explain the
very low efficiency (e.g., Melia \& Falcke 2001; Yuan, Quataert \&
Narayan 2003; Goldston, Quataert \& Tgumenshchev 2005; Liu \& Melia
2001; Liu, Melia \& Petrosian 2006).  The spectrum of Sgr A* has been
successfully modeled as a radiatively inefficient flow which is
carried across the event horizon before radiating away its energy
(Narayan et al.  1998).  More recently, detection of radio and
sub-millimeter polarization has allowed estimates of the integrated
electron density in the accretion region from $\sim$10 to 1000
Schwarzschild radii (Bower et al.  2003; Marrone et al.  2006).  These
estimates indicate much lower electron densities than predicted,
implying that perhaps most of the material is not reaching the central
black hole.  Theoretical work (e.g., Yuan et al.  2003; Sharma et al.\
2007) supports this picture, but leaves unanswered the question of
whether infalling material finds itself in a convective flow (Narayan
et al.  2002), a low-velocity outflow (Blandford \& Begelman 1999,
Igumenshchev et al.  2003), or a fast jet (Yuan, Markoff \& Falcke
2002).  The frequency-dependence of the size of Sgr A* in the radio
and sub-mm bands (e.g.\ Shen et al.\ 2005; Bower et al.\ 2006)
suggests that the plasma responsible for the emission at these
frequencies is not bound to the black hole but is escaping in a wind
(Loeb \& Waxman 2007).

Numerous multi-wavelength observations have studied Sgr A* in its
flaring state.  However, there are limited simultaneous observations
able to measure the correlation of the variability of Sgr A* in
different wavelength bands.  (e.g., Eckart et al.\ 2004, 2006;
Yusef-Zadeh et al.  2006a; Hornstein et al.  2007).  A variety of
mechanisms have been proposed to explain the origin of the variability
in the NIR and X-rays (e.g., Melia \& Falcke 2001; Yuan, Markoff \&
Falcke 2002; Liu \& Melia 2002; Yuan, Quataert \& Narayan 2003; Eckart
et al.  2004, 2006; Goldston, Quataert \& Igumenshchev 2005; Gillesen
et al.  2006; Liu et al.\ 2006a,b).  These studies imply that the NIR
is synchrotron emission while the X-rays are most likely produced by
synchrotron-self-Comption (SSC) or inverse Compton emission (Liu et al.  
2004).  Modelling of the
processes responsible for heating and cooling the transient population
of electrons responsible for the NIR and X-ray flares imply that it is
produced in localised regions in the inner part of the accretion flow
(Liu et al.  2006a,b; Bittner et al 2007).

The situation is less clear at sub-mm and radio frequencies where less
dramatic short-term variability coexists with longer term variations
presumably due to global changes in the accretion flow, and optical
depth effects likely play a role in moderating the short-term
variability (e.g. Goldston et al.  2005).  Previous simultaneous
near-IR and sub-mm measurements have shown two near-IR flares with
durations of about 20--35 min within 2.5 hours of an 850 $\mu$m flare
with an estimated duration of two hours (Yusef-Zadeh et al.  2006a).
The first near-IR flare appeared to coincide with the submm flare
whereas the second flare was delayed roughly by 160 minutes with
respect to the submm flare.  Two important points could be drawn:
First, assuming that near-IR and submm flares are related to each
other, this implies that the same synchrotron emitting electrons are
responsible for production of both near-IR and sub-mm flares.  The
second point is that it is not clear whether the sub-mm flare is
correlated simultaneously with the second bright near-IR flare, or is
produced by the first near-IR flare but with a time delay of roughly
160 minutes.  A delay between the peak emission in the near-IR and
sub-mm suggests that the enhanced synchrotron emission in the compact
region dominating the flaring emission is initially optically thick in
the sub-mm, becoming optically thin due to expansion (van der Laan 1966).

To follow up this idea, we studied the light curves of Sgr A* at 7 and
13 mm taken in 2005.  The light curves at both wavelengths show a
5-10\% increase of flux with respect to their quiescent levels.  We
found that the emission at 13 mm may lag that at 7 mm by $\sim$ 20
minutes, though with large uncertainty (Yusef-Zadeh et al.  2006b).
To confirm the time delay between flares at successively longer
wavelengths wavelengths and to search for a time lag or other
systematic correlation between other wavelength bands, we undertook a
series of detailed observations using the VLA , CSO and \Chandra
telescopes.

This paper discusses the results of observations made on 2006 July 17
when all three telescopes participated in observing Sgr A*
simultaneously.  The Keck and SMA also joined the 2006 July 17
observation, and their results will be described in an accompanying
paper (Marrone et al.  2007).  We also present the results of short
observations made on July 16, 2006 using the VLA and CSO. as well as
the cross correlation of flare emission at 7 and 13 mm observed with
the VLA on 2005 February 10 and 2006 February 10.  Finally, fitting a
simple adiabatically expanding plasma model gives reasonable matches
to the light curve at two or more frequencies placing strong
constraints on the physical parameters and size of the expanding
plasma that are consistent with the parameters inferred for the
accretion flow around Sgr A* and for the emission regions responsible
for the NIR flares (Liu et al.  2006a,b; Bittner et al.  2007).

\section{Radio, Submm and X-ray Observations}
\subsection{VLA Observations}

Using the Very Large Array (VLA) of the National Radio Astronomy
Observatory\footnote{The National Radio Astronomy Observatory is a
facility of the National Science Foundation, operated under a
cooperative agreement by Associated Universities, Inc.}, we carried
out observations in the B-array configuration on 2006 July 16 and 2006
July 17 for a total of 4 and 6 hours, respectively.  Details of the
total intensity and polarization calibration procedures observations
are identical to previous measurements, as described in Yusef-Zadeh et
al.  (2007), with the exception that these observations used two
sub-arrays simultaneously at 43 and 22 GHz.  Thus, the measurements
presented here are truly simultaneous in two observing bands.  Two
phase calibrators 1733-130 and 17444-31166 were employed.  The VLA is
undergoing an expansion and these observations used seven new Expanded
(EVLA) antennas and 20 original (VLA) antennas.  Because of higher and
more variable amplitude gains of the EVLA antennas than the VLA
antennas, we did not use any of the EVLA antennas in our analysis.
Thus, on the average a total of eight to ten antennas are used in each
sub-array.  The light curves at 43 and 22 GHz have used {\it
uv}-spacings longer than 100 k$\lambda$.  We discarded data from
antennas that showed large variation or high amplitude gains during
the run.  All antennas had similar gain amplitude during the entire
observation.  The 43 GHz data on 2006 July 16 were noisy and could not
be salvaged for time variability analysis.  The light curve of the
fast switching calibrator is obtained by calibrating it with the
second calibrator 1733-130.  Instrumental calibration was done in two
steps, one for the entire period of observation and the other on a
short time scale.  We only used 17444-31166 for polarization
calibration in both steps.  A more detailed account of short time
scale polarization calibration can be found in Yusef-Zadeh et al.
(2007).

Two other observations made on 2005 February 10 and 2006 February 10
used the BnA and A+Pt array configurations of the VLA to study the
time variability of Sgr A*.  Both observations used fast switching
techniques alternating between 43 and 22 GHz every few minutes.  Early
results including polarization results of these measurements are given
in Yusef-Zadeh et al.  (2006a, 2007).  These observations used the
same calibrators as in the July 2006 observations.
These radio observations are different  when compared 
with previous snapshot measurements (e.g., Zhao et al. 2004).   
First of all, the radio observations that have conducted in snapshot mode 
have no way of determining the flux stability of the complex gain 
calibrators.
The  44 and 23 GHz observations are most affected by
the opacity variability of the atmosphere.  Unlike previous  observations, 
we specifically used two nearby calibrators (one to independently calibrate
the other) in a fast switching mode (30 on calibrator and 90 seconds on 
source) in order to make sure that the change in the flux of Sgr A* is not 
due to the atmospheric effects.  These tests and techniques can not work 
when the data are taken in a snapshot mode.
Even if the calibration is done correctly in a snapshot observing 
mode, and the observed
variability is intrinsic to Sgr A*, it is not clear 
whether this is due to the change in the quiescent flux or true flaring
activity. Lastly, the lack of simultaneous coverage 
make it difficult to  claim time delays between peaks of emission in 
different  wavelength bands.  

\subsection{CSO Observations}

A $\sim3'$ field surrounding Sgr A* was observed at 850 \mic\ using
the Caltech Submillimeter Observatory and the SHARC II camera (Dowell
et al.  2003).  The observations on 2006 July 16 began at 5:36 UT and
ended at 10:50 UT, and the observations on 2006 July 17 began at 5:20
UT and ended at 10:21 UT. As was done in the past (Yusef-Zadeh al.
2006a), the field was observed with Lissajous scanning of the
telescope with an amplitude of $\sim$100\arcs\ and a period of
$\sim$20 seconds.  The scanned observing permits an accurate
measurement of all of the relative detector gains, and coverage of the
surrounding dust emission allows the atmospheric transmission to be
tracked better than using the CSO 225 GHz radiometer alone.

Absolute calibration was accomplished with observations of Callisto
(13.6 Jy assumed) and Neptune (27.5 Jy assumed), with an estimated
uncertainty of 10\%.  At the 20$''$ resolution of CSO at 850 \mic\,
the emission of Sgr A* is somewhat confused with the emission from the
surrounding molecular ring.  We estimate an absolute uncertainty of 1
Jy in the flux of Sgr A* due to confusion, while noting that the
accuracy of changes in flux is considerably better.

\subsection{Chandra Observations}

The imaging array of the Advanced CCD Imaging Spectrometer (ACIS-I;
Garmire et al.\ 2003) on-board the {\it Chandra X-ray Observatory}
(Wiesskopf et al.\ 2002) was used to observe \object{Sgr~A*}
simultaneously with the CSO and the VLA for 29.8~ks from 2006 July 17
04:15 to 12:37 (UT; ObsID \dataset[ADS/Sa.CXO#obs/6363]{6363}). 
ACIS-I was operated in timed exposure mode with detectors I0--3 and S2
turned on; the time between CCD frames was 3.141~s.  The event data
were telemetered in faint format.  The data were analyzed using the
\Chandra Interactive Analysis of Observations (CIAO) version 3.3.0.1
and the calibration database version
3.2.2\footnote{\url{http://cxc.harvard.edu/ciao}}.

Figure~1 shows the (net) {\it Chandra}/ACIS-I lightcurve of
\object{Sgr~A*} in the 2--8 keV band  obtained using a source aperture
radius of 1\farcs5 and a surrounding sky annulus extending from
2\arcsec\ to 4\arcsec, excluding regions around discrete sources and
bright structures (see Baganoff et al.  2003).  The bin size is
207.3~s.  For more details about the {\it \Chandra} data, see Marrone
et al.\ (2007).

\section{Results}

Figure 1 shows a composite light curve of flaring activity in the
X-ray (2-10 keV), 850$\mu$m, 7 and 13 mm bands on 2006 July 17.  The
most interesting result of this multi-wavelength campaign is the
detection of a strong X-ray and sub-mm flare.  The 43 and 22 GHz light
curves also show a relatively strong flare preceded by a weak flare
near 4.25 UT at 43 GHz.  The X-ray flare lasts for at least 90 minutes
with a sharp rise and a slow decay having a peak X-ray luminosity of
4$\times10^{34}$ ergs s$^{-1}$ (Hornstein et al.  2007).  The
asymmetric profile of the 850 $\mu$m flare shows a slower rise and
decay in its light curve than that of its X-ray counterpart.  The
duration of the submm flare is greater than 3.5 hours.  The
morphology, the duration and the time lag of the sub-mm flare with
respect to the bright X-ray flare can all be understood in the context
of expanding hot plasma (see $\S4$).  We also note a weak flare near
6h UT in the submm band with a peak flux of 200 mJy at 850 $\mu$m.
This weak flare was also detected at 1.2 mm based on simultaneous SMA
observations during this campaign (Marrone et al.  2007).
  
Simultaneous near-IR observations on 2006 July 17 also detected a
near-IR counterpart to the X-ray flare at K'-L' bands.  However,
near-IR observations started 36 minutes after the peak X-ray emission.
In spite of incomplete time coverage in near-IR wavelengths, Hornstein
et al.  (2007) argue that the near-IR and X-ray flare emissions are
associated with each other and simultaneous with no time delay.

\subsection{Cross Correlation Study}
\subsubsection{2006 July  17 and 2006 July 16}

Our cross-correlations use the Z-transformed discrete correlation
function algorithm (Alexander 1997; see also Edelson \& Krolik 1988).
This algorithm is particularly useful for analyzing sparse, unevenly
sampled light curves.  We identify the peak likelihood value, and a
1-$\sigma$ confidence interval around that value, using a maximum
likelihood calculation (Alexander 1997).

The top panel of Figure 2a shows the light curves of Sgr A* at
850$\mu$m and X-rays; the corresponding cross correlation peak is
displayed in the bottom panel.  Assuming that the two flares are
related to each other, the cross correlation plot suggests that the
sub-mm peak lags the X-ray peak by 110$\pm17$ minutes.  Also, assuming
that the X-ray flare has a simultaneous near-IR counterpart, as
Hornstein et al.  (2007) suggest, the 110m time delay seen in Figure
2a is consistent with that observed between the near-IR and sub-mm
peaks measured on 2004 September 4 (Yusef-Zadeh et al.  2006a).  These
time delay measurements imply that submm flare emission at 850 $\mu$m
is optically thick.

Figure 2b shows a cross-correlation peak between 7 and 13 mm emissions
for the 2006 July 17 data.  The maximum likelihood peak is
20.4$\pm6.8$ minutes.  The time lag between the 7 and 13 mm peak flare
emission, as shown in Figure 2b, is consistent with a 20 minute time
delay at these frequencies reported earlier (Yusef-Zadeh et al.
2006b).

We also performed a cross correlation analysis between the radio and
submm flare emission of 2006 July 17 (Figure 3a).  However, the strong
flare emission in radio wavelengths occurs before the strong submm
peak seen near 7:30h UT. There is a weak submm flare that is evident
near 6h UT. This weak submm peak, which is clearly visible in the SMA
light curve of 2006 July 17, has a better signal-to-noise at 1.3mm
than at 850$\mu$m (Marrone et al.  2007).  Due to the long duration
and relatively high frequency of the radio and submm flares, as well
as the lack of simultaneous temporal coverage between VLA and CSO
observations, the association of flares in radio and submm wavelengths
is not clear.  One possibility is that the strong 7 mm flare that
peaks near 6:30-7h UT, as seen in Figure 1, is associated with the
strong submm flare that peaks near 7:30h UT. The strong peak in the
cross correlation plot in Figure 3a shows a negative time lag of
$\sim1$ hour for these two strong flares in radio and submm
wavelengths.  This result, however, implies that the strong submm peak
is delayed with respect to the strong 7 mm peak by 66$^{+18}_{-1.8}$
minutes.  This is, of course, not consistent with the plasmon model of
expanding plasma which predicts that low frequencies are delayed
relative to high frequencies.  On the other hand, Figure 3a also shows
a weaker cross correlation peak with a positive time lag.  Thus, we
infer that the faint submm flare and quiescent emission with a peak
flux of 2.7 Jy centered around 6h UT is likely to be associated with
the strong 7 mm flare and quiescent emission with a peak flux of 1.7
Jy.  This also suggests that both the X-ray and the strong submm
flares are not related to the radio flares observed at 7 and 13mm.
Given the uncertainly due to incomplete time coverage, the time lag
between the weak 850 $\mu$m and 7 mm flares is estimated to be greater
than 1.25 hours.

We also investigated the cross correlation of 850$\mu$m data with 13
mm data taken on 2006 July 16.  Figure 3b shows enhanced submm
emission between 6:45h UT and 8:0h UT. Radio emission at 13 mm also
shows the beginning of a flare between 7:30 and 8h UT. Again, the lack
of simultaneous temporal coverage between VLA and CSO observations
makes the cross correlation analysis quite uncertain.  However, a peak
is noted with a positive time lag in the bottom panel of Figure 3b.
The delay between 850 $\mu$m and 13 mm is estimated to be roughly
65$^{+10}_{-23}$ minutes.  This time delay is consistent with the time
lag noted with the 2006 July 17 cross correlation data, as shown in
Figure 3a.

\subsubsection{2006 February 10 and 2005 February 10 }

We applied the cross correlation analysis to the 7 and 13 mm data
taken on 2006 February 10 and on 2005 February 10 in the A-array plus
Pie Town and the BnA hybrid configurations of the VLA, respectively
(Yusef-Zadeh et al.  2006a, 2007).  Figure 4a,b shows the cross
correlation plots of 7 and 13 mm data corresponding to these
observations.  The 7 and 13 mm light curves from both observations are
made based on {\it uv} data greater than 100 k$\lambda$.  The maximum
likelihood values with one $\sigma$ error bar are 30$\pm12$ and
20$\pm6$ minutes for the 2006 February 10 and 2005 February 10 data,
respectively.  In both measurements, the strong peak indicates clearly
that the 13 mm peak lags behind the 7 mm peak flare emission.  Table 1
gives a summary of the time lag measured from cross correlating radio,
submm and X-ray data sets.

\subsection{43 and 22 GHz Polarization}

The four panels of Figure 5a,b show the variation of the polarization
plots at 7 and 13 mm, respectively, taken on 2006 July 17.  The top
three panels show the Stokes Q, U, and I whereas the bottom two panels
show the degree of polarization and polarization angle.  The data are
binned every 20 minutes at both 7 and 13 mm.  The average degree of
polarization at 7 and 13 mm are 1$\pm0.2$ \% and 0.7$\pm0.1$\%,
respectively. 

We also obtained light curves of Stokes Q, U and I using 
1733-130 instead
of 17444-31166. We could not confirm the light curves of Stokes Q and U,
as shown in Figure 5. We believe the reason for such a discrepancy is due
to the residual instrumental errors of the calibrators as a function of
elevation. If the residual errors are not taken out properly, they create
different polarization light curves. We believe that 17444-31166 is a
better polarization calibrator for this purpose as was observed every 90
seconds while tracking Sgr A* within 2$^0$ of its position. On the other
hand, 1733-130 is observed once every 20minutes and is 16$^0$ away from
from Sgr A*.  It is possible that the stress due to gravity is responsible
for not accounting properly for instrumental polarization correction as a
function of time. Future observations with additional polarization
calibrators are needed to confirm the low level of polarization we have
reported here.

We note that both the degree of polarization and polarization angle
tend to change during the flaring state when compared with those
during the quiescent phase of Sgr A*.  In particular, we note a
possible increase in the degree of polarization to 2.5\% and 1.5\% at
7 and 13 mm, respectively.  The polarization angle during the
quiescent phase is close to 180 or 0 degrees at 7 and 13 mm, but a
variation of about 100 degrees is detected during the flaring phase at
both 7 and 13 mm.  The average values of polarization angle at 7 and
13 mm during the flaring phase (i.e. $> 6$h UT) are estimated to be
106 and 87 degrees.  The errors in the polarization angle for
individual data points are obtained by taking the nominal Stokes Q and
U values plus the derived 1-$\sigma$ errors in the respective Stokes
parameter to determine one limit in the polarization angle.  The
nominal Q and U values minus the 1-$\sigma$errors are used to
determine the other limit relative to the nominal polarization.  The
polarization angle measurements correspond to a RM of $-2.5\times10^4$
rad m$^{-2}$ with no phase wrap; the RM estimate assumes the quadratic
relationship between RM and the observed frequency.  However, recent
analysis of polarized emission from Sgr A* suggest that relativistic
RM of polarized radiation should be considered (Ballantyne, Ozel, \&
Psaltis 2007; Yusef-Zadeh et al.  2007).

\section{Modeling of sub-mm and radio flares}

Models for the variable emission from Sgr A* have primarily focussed
on the relationship between the dramatic flaring observed in the NIR
and X-ray bands (Baganoff et al.  2003, Eckart et al.\ 2004, 2006;
Yusef-Zadeh et al.  2006a; Hornstein et al.  2007).  In these models
the IR emission is produced by synchrotron emission from a transient
population of near-Gev electrons in a $\sim$10--100\,G magnetic field,
with the x-rays produced either by synchrotron from higher energy
electrons (Yuan et al.  2003) or by inverse Compton scattering of
either the sub-mm from the surroundings or directly associated with
the transient population (ie synchrotron-self-Compton) (e.g. Yuan et
al.  2003, 2004; Liu et al.  2004, 2006a,b).  Scattering of NIR
photons by the electrons responsible for the quiescent sub-mm emission
may also contribute to the X-ray flux (Yusef-Zadeh et al.  2006).
These models constrain the size of the emitting region to be a few
Schwarzschild radii or less (Liu et al.  2004).  The transient
population of accelerated electrons may be produced via reconnection,
acceleration in weak shocks (e.g. Yuan et al.  2003) or by heating by
plasma waves (Liu et al.  2006a,b) driven by instabilities in the
accretion flow or dissipation of magnetic turbulence.  Detailed
modelling of the latter process implies that the regions are compact
suggesting that the flaring may be used to study the detailed
evolution of the accelerated electron population in response to
heating and radiative mechanisms (Bittner et al.  2007).

At lower frequencies, successive delays have been tentatively detected
between flares at 350, 43 and 22 GHz (Yusef-Zadeh et al.\ 2006a),
behaviour that is reminiscent of the emission from an adiabatically
expanding synchrotron source that is initially optically thick, and
becomes optically thin at successively lower frequencies as the source
expands (cf.\ van der Laan 1966).  This may reflect (for example) the
expansion of a buoyant blob of synchrotron-emitting plasma created in
a corona by a disk instability or a bright emission spot advected by an
expanding wind in the vicinity of the black hole.  Here we adopt a
phenomenological approach focussed on the empirical behavior of the
flares that is divorced from the uncertain global structure and 
dynamics of the outflow.

It turns out that this model matches the shapes of the profiles and
the delay between frequencies surprisingly well, and that the derived
emission region size, density and magnetic field steength are in broad
agreement with the conditions inferred in global models of the
accretion flow and in the detailed models for flaring in the NIR
(Bittner et al.\ 2007).

Our earlier phenomenological modeling of the time delay assumed that
at 350 GHz the flare emission was initially optically thin and showed
that a steep energy spectrum ($\sim$E$^{-3}$) could explain the
relative flux of the peak emission at lower, initially optically thick
frequencies.  Here we futher test this model by applying it to the
delayed flaring seen in the curent observations.  The presence of a
possible 110-minute time delay between the peaks of X-ray and 850
$\mu$m light curves suggests that the sub-mm flare emission may, in
this case, be optically thick.  Thus, here we investigate optically
thick sub-mm and radio emission in the context of adiabatic expansion
of hot plasma (van der Laan 1966) using different energy spectra of
nonthermal particles.  We also determine the physical parameters of
the expanding blob models which then are used to model the
polarization characteristics of the observed light curves in radio and
submm wavelengths.

\subsection{Time Delay as a Function of Spectral Index}

The phenomenological model assumes that the flaring emission at sub-mm
to radio frequencies is dominated by synchrotron emission from a 
compact
expanding region filled with relativistic electrons with an energy spectrum
$n(E) \propto E^{-p}$ threaded by a magnetic field.  For simplicity,
the particle density and spectrum and magnetic field strength are
taken to be uniform.  As the emitting region expands the magnetic
field declines as $R^{-2}$ because of flux-freezing, the energy of
each relativistic particle declines as $R^{-1}$ because of adiabatic
cooling and the energy-integrated density of relativistic particles
scales as $R^{-3}$.  Following van der Laan (1966), the
flux density scales as
\begin{equation}
    S_\nu = S_0 \left(\frac{\nu}{\nu_0}\right)^{5/2} 
    \left(\frac{R}{R_0}\right)^3 
    \frac{1-\exp(-\tau_\nu)}{1-\exp(-\tau_0)} \,,
    \label{eq:Snu}
\end{equation}
where the synchrotron optical depth at frequency $\nu$ scales as
\begin{equation}
    \tau_\nu = \tau_0 \left(\frac{\nu}{\nu_0}\right)^{-(p+4)/2}
    \left(\frac{R}{R_0}\right)^{-(2p+3)}\,.
    \label{eq:taunu}
\end{equation}
Here $\tau_0$, the optical depth at which the flux density for any
particular frequency peaks, depends only on $p$ through the condition
\begin{equation}
    e^{\tau_0} - (2p/3 + 1)\tau_0 - 1 = 0
    \label{eq:tau0}
\end{equation}
and ranges from 1 to 1.9 as $p$ ranges from 1 to 3 (Yusef-Zadeh et
al.\ 2006b).  Thus given the particle energy spectral index $p$ and
the peak flux $S_0$ in the flaring component of the light curve at a
reference frequency $\nu_0$, this model predicts the variation in flux
density at any other frequency.  A model for $R(t)$ is required to
convert the dependence on radius to time: as a first approximation we
adopt a simple linear expansion at constant speed $v$, so that $R =
R_0 + v\,(t-t_0)$ where $t_0$ is the time of the peak flux at frequency
$\nu_0$.

By way of illustration, we choose a peak flux $S_0 = 1.5$\,Jy at
$\nu_0 = 350$\,GHz, consistent with the observations
presented here.  Figure 6 shows the resulting light curves at
sub-mm and radio frequencies for particle energy indices $p=2$ and 0.5.
This behaviour reflects the combined effects of adiabatic cooling and
declining magnetic field, which imply that the {\it initial} energy of
the electrons responsible for the peak emission increases as the
plasma evolves.  A flatter spectrum implies a relatively larger number
of high energy particles at t=0, and therefore a greater expansion is
required to cool them.  The consequence of flattening the energy
spectrum of particles is therefore to increase the time delay between peaks at
various frequencies and to lengthen their durations.

\subsection{Modeling the Light Curves}

For a single frequency the functional form for a flare
at one frequency uses one more parameter than a Gaussian fit (i.e.\
$p$ and the expansion speed $v$, instead of the Gaussian FWHM), so
that one expects a better fit than would be obtained with a gaussian,
regardless of the merits of the model.  Fortunately the
contemporaneous light curves at other frequencies -- including the
delay time -- are then fully determined without additional parameters.
Therefore simultaneous light curves at two or more frequencies form a
good test of the model.

The observables measured from a typical light curve at frequency
$\nu_0$ are the timing and magnitude of the peak flux ($t_0$, and
$S_0$), the particle index $p$ and expansion speed in units of $R_0$
per unit time (which together determine the asymmetry and width of the
flare), and the background level.  The most uncertain of these is the
value of the background, i.e. the quiescent flux, as this may slowly
vary over the course of a flare.

The physical parameters of the emitting region are then determined as
follows.  The index $p$ determines $\tau_0$, and so the value of $S_0$
fixes the radius $R_0$ via $S_0 = \pi R_0^2 / d^2 \times
(1-\exp(-\tau_0))$ where d is the distance to the Galactic center.
Then the expansion speed is determined in physical units.  The
determination of the magnetic field and electron density from the
optical depth and size of the emitting region relies on assuming
minimum and maximum energies (we adopt E$_{min} = 1 $ MeV and E$_{max}
= 100$ MeV, respectively) and equipartition between the magnetic field
and relativistic electrons.

We do not perform maximum likelihood fitting given the
phenomenological nature of the model, being content to achieve
reasonable matches with the observed light curves.  We modeled the
sub-mm and radio light curves on 2006 July 17 and the radio data on
2005 February 10 and 2006 February 10.  The model parameters are
presented in Table 2.

The model for the submm light curve of 2006 July 17 is plotted as a
solid line in Figure 7a.  The sub-mm flare shows a slow decay and its
asymmetric profile is fitted with a peak (excess) flux of 0.82 Jy at
850$\mu$m (350 GHz) at 7.65h UT. The particle spectral index p=1 (or
$\alpha$=1), an expansion speed of v = 0.0028c and a magnetic field of
B = 76 G are derived from the fit.  Note that the synchrotron cooling
time for electrons emitting ar 350 GHz is $\sim45$ minutes, so that
this may play a role in determining the decay of the light curve.

Separately from the submm data, the light curves of the radio flare
that peak prior to the rise of the submm flare on 2006 July 17 are
jointly fitted, as shown in Figure 7b.  The peak flux of 0.23 Jy at 7
mm (43 GHz) at 6.55h UT is used to derive the value p=1, v = 0.12c and
a magnetic field of 11 G, as shown in Figure 7b.  The quiescent flux
was estimated to be 1 and 1.46 Jy at 22 and 43 GHz, respectively, with
a spectral index of 0.58.

The model light curves of the 2005 February 10 data at 43 and 22 GHz
are shown in Figure 8a.  The derived quantities are p=1.5, v=0.018c
and B=12 G. We also derived the physical parameters of the flare
emission for the 2006 February 10 observation.  The light curve of
this observation shows two overlapping flares on top of a fairly
strong quiescent emission when compared to the flare emission observed
on 2005 February 10.  The fit to the light curve is shown in Figure
8b.  The model uses peak flux of 300 and 240 mJy at 43 GHz
centered at 14.3h and 16.3h UT for the first and second flare,
respectively.  The derived physical parameters for both flares are
quite similar to each other (see Table 2) except that the value of
particle energy index of the first flare (p=1) is flatter than that of
the second flare (p=2).

\subsection{Theoretical Polarization Curves}

The parameters derived from modelling the variation in intensity can
be used to compute the polarization properties of the transient
emitting region, in other words to produce light curves in all four
Stokes parameters I,Q, U and V. To illustrate this we compute the
variations in Q and U for the physical parameters inferred of the
emitting region inferred for the 2006 July 17 flare.  To do this we
use the analytic (but complicated!)  solutions for the Stokes vector
emerging from a homogeneous synchrotron-emitting sphere threaded by a
uniform magnetic field given by Jones \& O'Dell (1977).

The predicted polarization is sensitive to the assumption that the
magnetic field in the emitting region is constant, as any spatial
variation in B will tend to reduce the amplitude of the polarization.
Indeed, we find that the model overpredicts the degree of polarization
by a factor of around three, so for the purposes of illustration we
reduce $U$ and $Q$ by this factor.  Two additional parameters are
needed to account for the orientation of the magnetic field to the
line of sight: we find a reasonable match to U and Q at 43 GHz for a
field lying in the plane of the sky with PA= 20\deg.  The quiescent
emission at 43 GHz is linearly polarized with $Q=-4$\,mJy and
$U=6.6$\,mJy, so we add these amounts to the model curves.

Figure 9 shows the fit to all Stokes parameters of the 43 and 22 GHz
light curve of the 2006 July 17 flare.  The match at 22 GHz is less
compelling but despite the uncertainties, both in the data and the
model, these results indicate the potential for improved measurements
of polarization at multiple frequencies to severely test the emission 
model.

\section{Discussion}

The particle acceleration mechanism responsible for the NIR flaring
associated with Sgr~A* is not yet conclusively identified.
Possibilities include acceleration in weak shocks (e.g. Yuan et al.
2003), magnetic reconnection events in a disc corona (e.g. Yuan et al.
2004) or acceleration by a spectrum of plasma waves associated with
the dissipation of maagnetically-driven turbulence (Liu et al.  2004;
2006a,b).  Modelling of the NIR and X-ray flaring implies that the
emission regions are compact, so that the flaring can be used to study
the detailed evolution of the accelerated electron population in
response to time-dependent heating and radiative mechanisms (Liu et
al.  2006a,b; Bittner et al.  2007).

At the frequencies of interest here -- below 350 GHz -- we cannot
directly investigate the mechanism responsible for the energization of
the electrons because the flaring emission is optically thick and the
synchrotron cooling time of the emitting particles is longer than the
flare duration.  However, the decaying part of the light curves are
sensitive to the electron spectrum over a range of energies because a
weakening magnetic field means that the energy of the electrons
responsible for the synchrotron emission at a particular frequency
increases with time; in addition adiabatic cooling of the particles
implies that the electrons had an even higher energy at the onset of
the flare.  Adiabatic cooling in the absence of significant
synchrotron losses implies that the energy of a particular electron
scales as $1/R$, and the magnetic field strength scales as $R^{-2}$.
Thus the energy of the electrons largely responsible for synchrotron
emission at a particular frequency scales as $R$ and their energy when
the population was created scales as $R^{-2}$.  Thus modelling of the
flare profiles potentially tells us something of the initial
accelerated population over about a factor of 5--10 in energy.

In our modelling we assumed for simplicity a power-law spectrum
($E^{-p}$) and inferred $p=1$--2 for the asymmetric and more symmetric
profiles of light curves, respectively.  Fermi acceleration by a
single adiabatic shock produces a power law energy distribution with
p$\sim$2 (e.g., Blandford and Eichler 1987).  A harder particle
spectrum can be produced if relativistic particles carry away a
substantial fraction of the kinetic power of the shock (Drury \&
V\"olk 1981), or towards high energies by acceleration in multiple
shocks (Melrose and Pope 1993; Pope and Melrose 1994).  Note, however,
that many plausible models predict a relativistic Maxwellian by
competition between accelerative and radiative processes (e.g.\
Bittner et al.  2007).  It would therefore be of interest to compute
the flaring at low frequencies assuming such spectra, but we leave
that exercise to the future, noting that our inferred value of $p$ may
reflect the slope of the electron spectrum over a restricted range of
energies.

The variation of energy index of particles may also have implications
on the origin of the flaring component of X-ray emission.  It has been
argued previously that the X-ray counterparts to the near-IR flares
are unlikely to be produced by synchrotron radiation in the typical
$\sim10$\,G magnetic field for two reasons.  First, emission at
10\,keV would be produced by 100\,GeV electrons, which have a
synchrotron loss time of only 20\,seconds, whereas individual X-ray
flares rise and decay on much longer time scales (Baganoff et al.
2001).  Second, the observed spectral index of the X-ray flares,
$S_\nu \propto \nu^{-0.6}$ (Belanger et al.  2005), does not match the
near-IR to X-ray spectral index of $-1.35\pm0.2$ (Eckart et al.
2006).  We favored an inverse Compton model for the X-ray emission,
which naturally produces a strong correlation with the near-IR flares.
In this picture, sub-millimeter photons are up-scattered to X-ray
energies by the electrons responsible for the near-IR synchrotron
radiation.  In the inverse Compton scattering (ICS) picture, the
spectral index of the near-IR flare must match that of the X-ray
counterpart.  The spectral indices of the X-ray and of the decaying
part of the near-IR light curve on 2006 July 17 are estimated to be
$\alpha = -0.3\pm0.6$ and -0.6$\pm0.2$, respectively, where F$_{\nu}
\propto \nu^{\alpha}$ (Hornstein et al.  2007).  These spectral
indices are consistent within errors with the ICS picture.  Hornstein
et al.  (2007) report a peak X-ray flux of 4$\times10^{34}$ ergs
s$^{-1}$.  In the ICS picture, the peak flux at K-band is estimated to
be $\sim$45 mJy assuming that the magnetic field is B=20G and the
near-IR spectral index $\alpha$=-0.5.  The estimated near-IR flux
depends on the submm source size and is assumed to have a radius of
3R$_s$, which is quite uncertain.

The emission from the innermost region of Sgr A* can be divided into
quiescent and flaring components.  The quiescent flux dominates the
emission at low energies (e.g., radio and submm wavelengths) whereas
the flaring emission dominates in near-IR and X-ray wavelengths.
Although a number of models have been proposed to explain the low
luminosity of the quiescent flux and the origin of the flaring
activity, the relationship between the quiescent and flaring states of
Sgr A* is not fully understood.  It is instructive to examine the
relationship between these two components, The light curves shown in
Figures 2 and 3 indicate clearly that the quiescent emission at 43 and
22 GHz varies on different days.  The most dramatic increase of about
30\% from 2005 February 10 to 2006 February 10 is noted.  Figure 3
shows the quiescent emission of 2006, February 10 ranging between 2.2
and 2.3 Jy at 43 GHz whereas the light curve of Sgr A* a year earlier
on 2005 February 10, shows a flux of $\sim 1.65$ Jy at the same
frequency.  Similar percentage daily changes can also be noted at
22 GHz.  At sub-mm band, the 850 $\mu$m (350 GHz) mean flux dropped by
1.5 Jy between 2005 July/August and 2006 July, This corresponds to a
drop of 50\% of the quiescent flux in 2006 July with a very high
statistical error.  This suggests that the emission at submm and radio
wavelengths, like those in near-IR and X-rays could be mainly due to
flaring activity and there is no true quiescent flux in the broad band
spectrum of Sgr A*.

Thus, it appears that some similarities that can be drawn from
comparing the broad band and flaring spectra of the quiescent flux of
Sgr A*.  One is the fact that sub-mm and radio emission is optically
thick in both regimes.  The peaks of both the broad band quiescent and
flare spectra turnover somewhere in the sub-mm band.  In addition, the
fact that both quiescent and flare components vary in the optically
and optically thin regions gives additional support to the idea that
much of the quiescent emission could be due to flaring events.
Near-IR light curves show that Sgr A* is active as much as 30\% of the
time.  In the context of expanding hot plasma blobs, the variation of
flux at optically thick submm and radio wavelengths may indeed be
responsible for varying quiescent flux of Sgr A*.  A more detailed
study of the relationship between flaring and quiescent flux of Sgr A*
should explore the possibility that much of the quiescent emission may
be generated as a result of a sea of simultaneous low amplitude flare
emission.

\section{Concluding Remarks}
In summary, the results presented here confirm the picture of
expanding synchrotron emitting plasma as a number of observations in
radio show evidence of time delay between the peaks of flare emission.
The first evidence of a time delay between X-rays and 850 $\mu$m
emission provides additional support for the optically thin
near-IR/X-ray flare emission leading the optically thick sub-mm and
radio flares by about 2-3 hours.  In fact, this time delay clarified
earlier simultaneous near-IR and sub-mm observations in 2004 that
reported sub-mm flare lagging a near-IR flare by 160 minutes
(Yusef-Zadeh et al.  2006a).  The observed duration and the time delay
of optically thick flare emission is consistent with the 
model.

The modeling of the light curves has given us a handle on a number of
physical quantities associated with flaring activity including the
initial magnetic field, the energy spectrum of the particles and the
expansion speed, the magnetic field with a strength of 76 G derived
from fitting optically thick submm light curve of 2006 July 17 is
relatively high compared to the estimate of the magnetic field from
model fitting of radio data.  This high magnetic field corresponds to
a synchrotron cooling time of 45 minutes in submm band, comparable to
the decay of of the submm flares.  This partially undermines the
argument that dynamical cooling of submm flares must be important,
because of the long synchrotron lifetime of submm emitting electrons.
Obviously, future correlation of submm and radio light curves are
critical to examine the  model and measure the initial magnetic
field of the expanding plasma.

The suggestion that the emitting regions responsible for the flaring in the
sub-mm and radio are undergoing expansion is consistent with the idea of a
significant large scale outflow of the material associated with the sub-mm
radio quiescent spectrum of Sgr A* (e.g. Loeb \& Waxman 2007).  Note,
however, that our inferred expansion speeds need not correspond to the bulk
velocity in the flow.  Assuming that the bulk motion is large, then there
may be some signature of outflowing material from Sgr A* although it is not
clear if the outflowing material will be collimated as there is no
definitive evidence that Sgr A* has a disk.  Indeed, there are some hints
that a collimated outflow may arisen from Sgr A* in the past.  For example,
there is the intriguing radio continuum image of the inner few arcseconds
of Sgr A* (Yusef-Zadeh, Morris \& Ekers 1992).  We speculate that these
large-scale (10$^{17}$ cm) blob-like structures could potentially be
associated with a past outflow activity from Sgr A*.

\acknowledgements
We are grateful to Tal Alexander for the use of his ZDCF code.  COH
acknowledges the support of the Lindheimer Postdoctoral Fellowship.
Research at the CSO is supported by the NSF under contract AST
0540882.  F.~K.~B.\ received support for this work from NASA through
\Chandra Award Numbers G05-6093X and G06-7041X, issued by the \Chandra
X-ray Center under contract NAS8-03060, and SAO Award Number
2834-MIT-SAO-4018.

\newcommand\refitem{\bibitem[]{}}


\begin{figure}
\includegraphics[scale=0.8,angle=0]{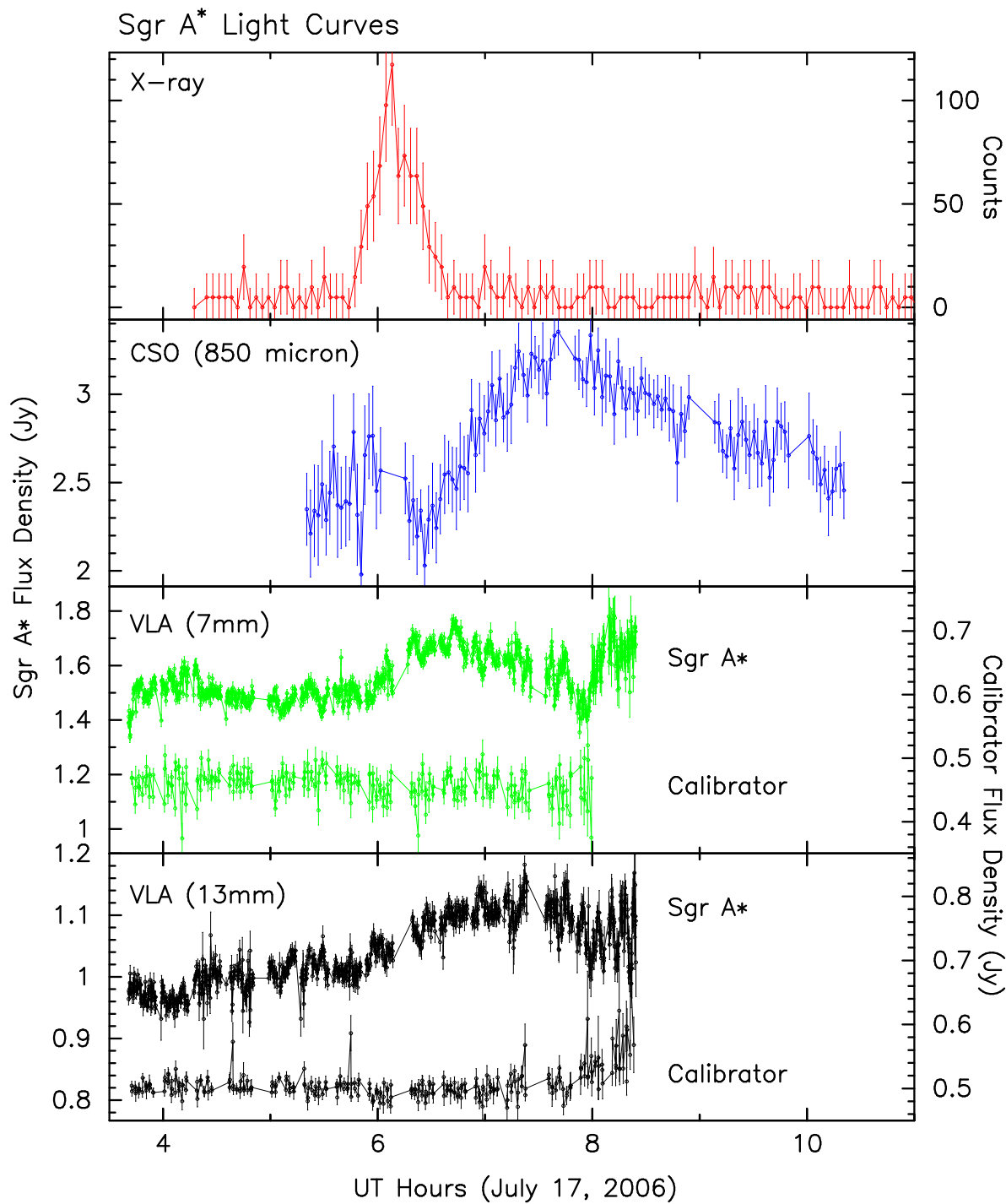}
\caption{{} The light curves of Sgr A* 
is shown during the 2006 July 17  observing campaign 
in four different panels using \Chandra, CSO and VLA observatories. 
The fluxes are measured in Jy except for the X-ray data which are  shown as a count rate. 
The  X-ray (2-8 keV), submm (850 $\mu$m) and  radio (7 and 13 mm) data are   binned 
every 207 seconds, 20 minutes and  9 seconds,  respectively. The light curves of 
the phase calibrator 17444-31166 
are  also presented at 7 and 13 mm  in  the bottom two panels. Note that the 
flux increase at 7mm near 8.2h UT is likely to be due atmospheric errors at
low elevation.}
\end{figure}  

\begin{figure}
 \includegraphics[scale=0.4,angle=0]{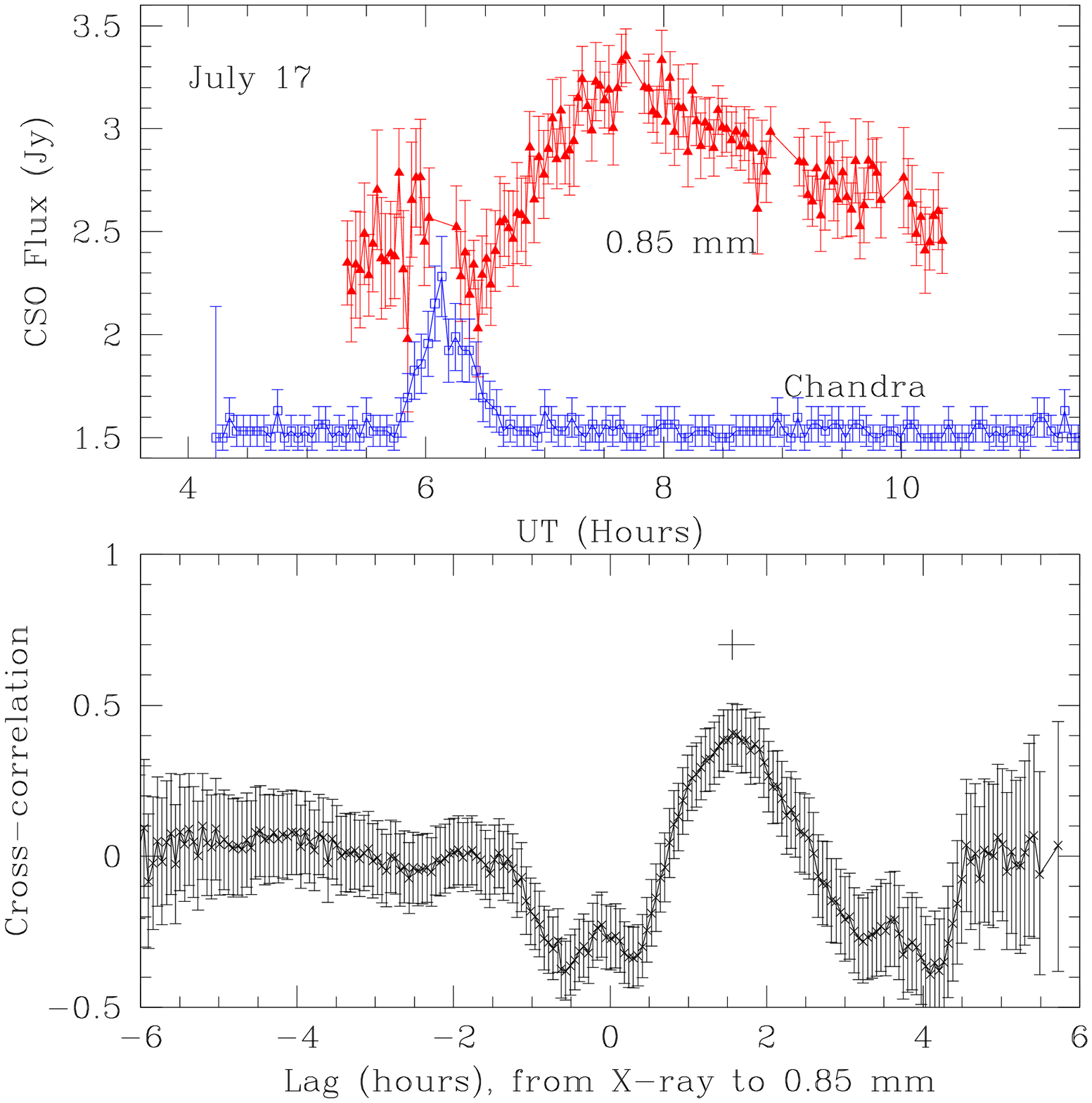}
  \includegraphics[scale=0.4,angle=0]{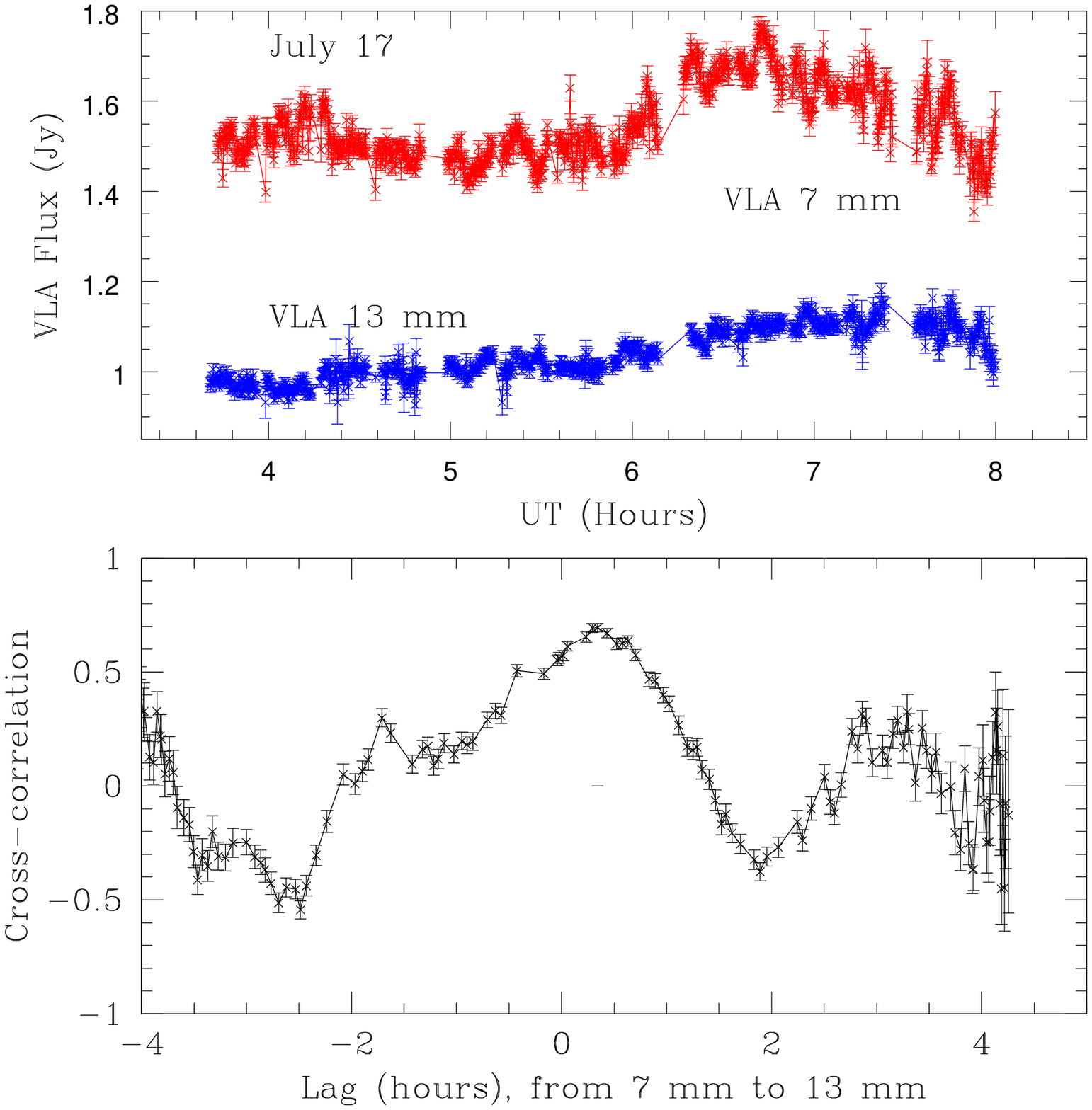}
\caption{{\bf (a) Left} the light curves of X-ray and submm emission and the corresponding 
cross correlation plot showing a time lag 
110$\pm17$ minutes. 
{\bf (b) Right} The cross correlation and light curves of 7 and 13 mm
 emission with a time lag 
of 20.4$\pm6.8$ minutes. The  
maximum  likelihood peak with a 1-$\sigma$ error bar,  
is  shown in the bottom panels of each figure. 
}
\end{figure}

\begin{figure}
 \includegraphics[scale=0.54,angle=0]{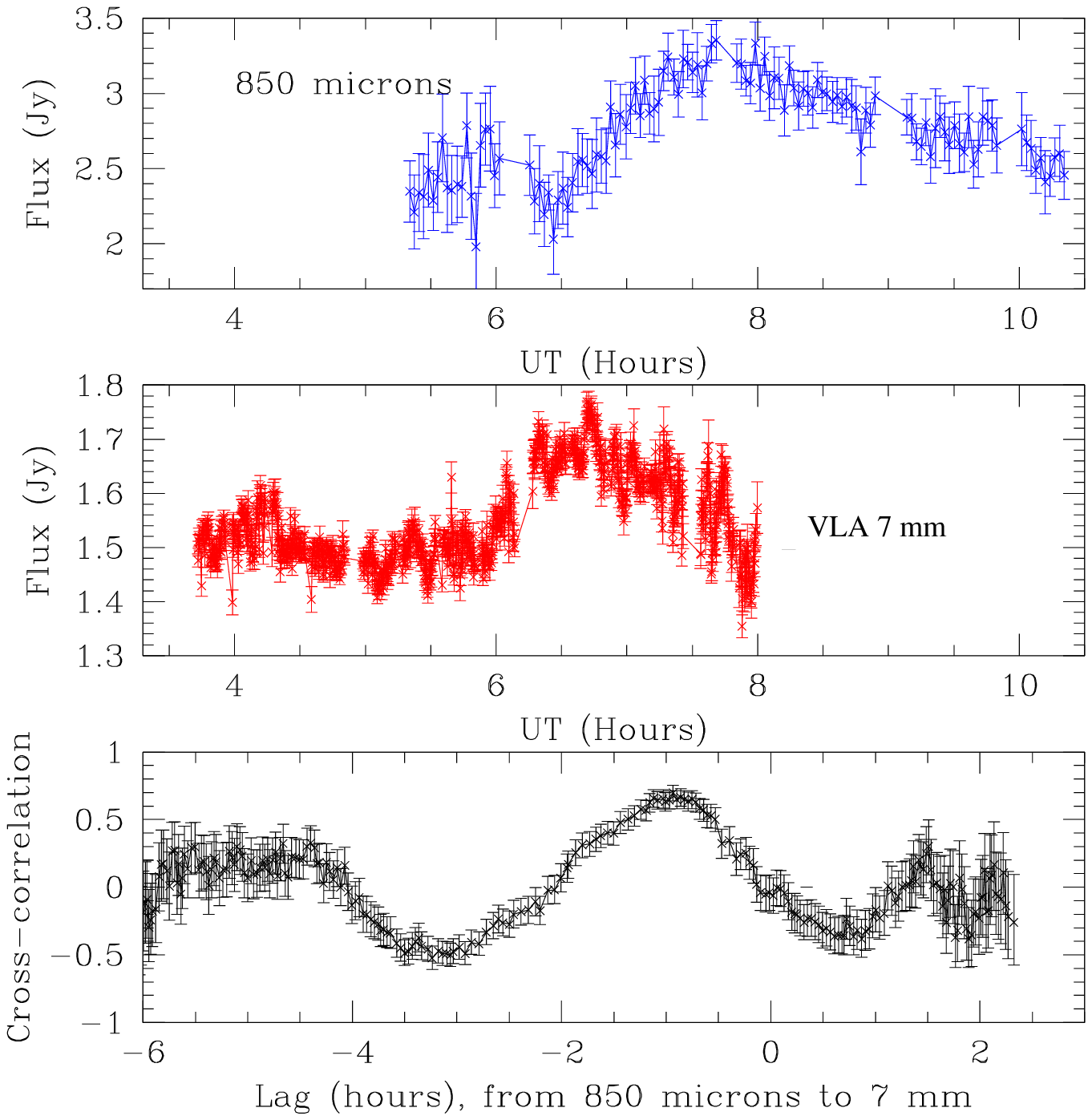}
  \includegraphics[scale=0.4,angle=0]{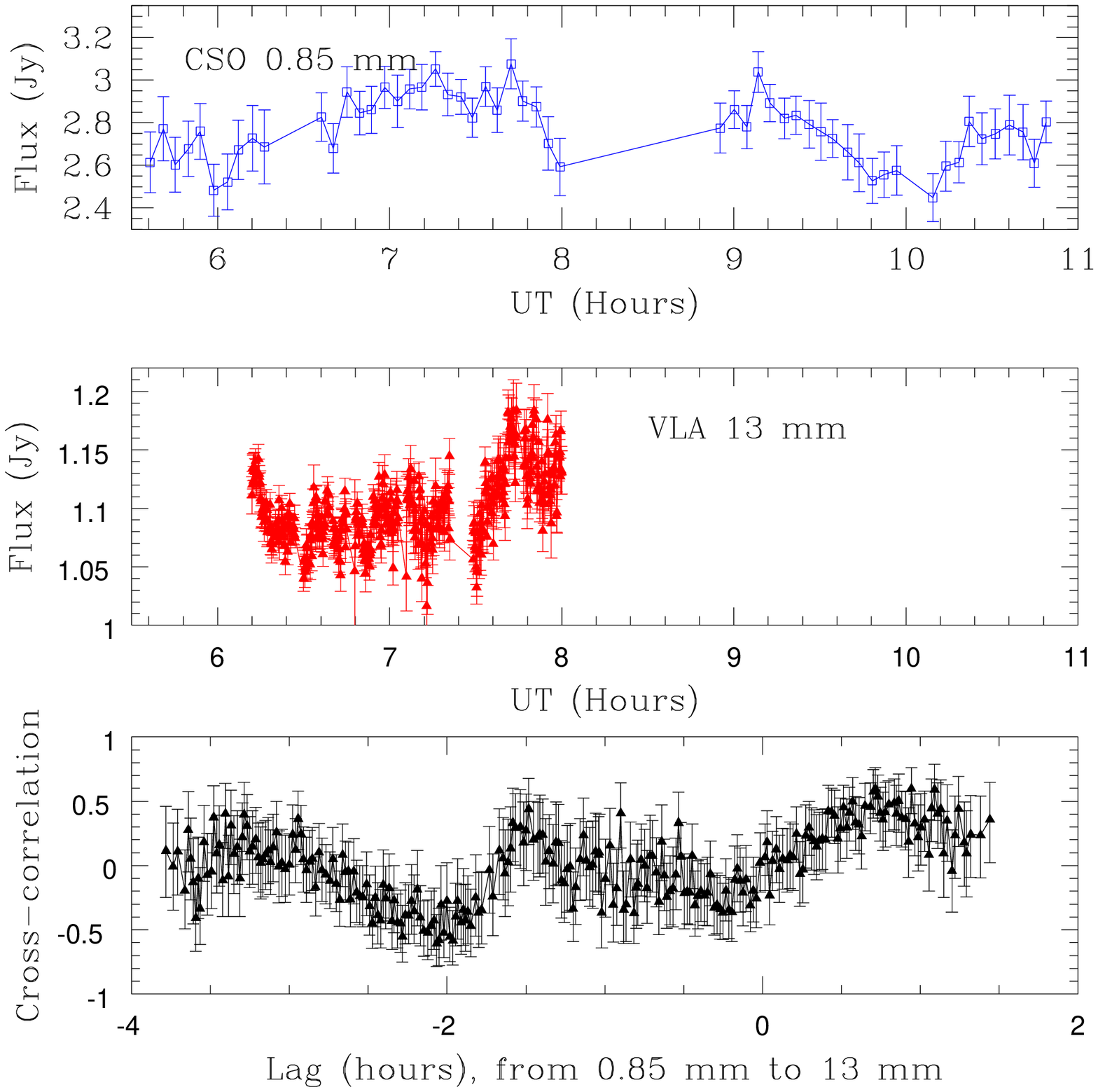}
\caption{{\bf (a) Left} Light curves of 850$\mu$m (blue color) and 7 mm (red color) 
emission 
on 2006 July 17  and the 
corresponding 
cross correlation plot. 
The cross correlation plot shows radio emission peak lags behind 
the  weak submm peak emission near 6h UT by  1.25 hours. 
{\bf (b) Right} 
 Light curves of 850$\mu$m or 0.85 mm (blue color)  and 13 mm (red color) flare  emission 
on July 
16, 2006 and the corresponding cross correlation plot. 
Both plots show that the peak  radio emission lags behind 
the submm peak by  45$\pm30$m. }
\end{figure}

\begin{figure}
 \includegraphics[scale=0.4,angle=0]{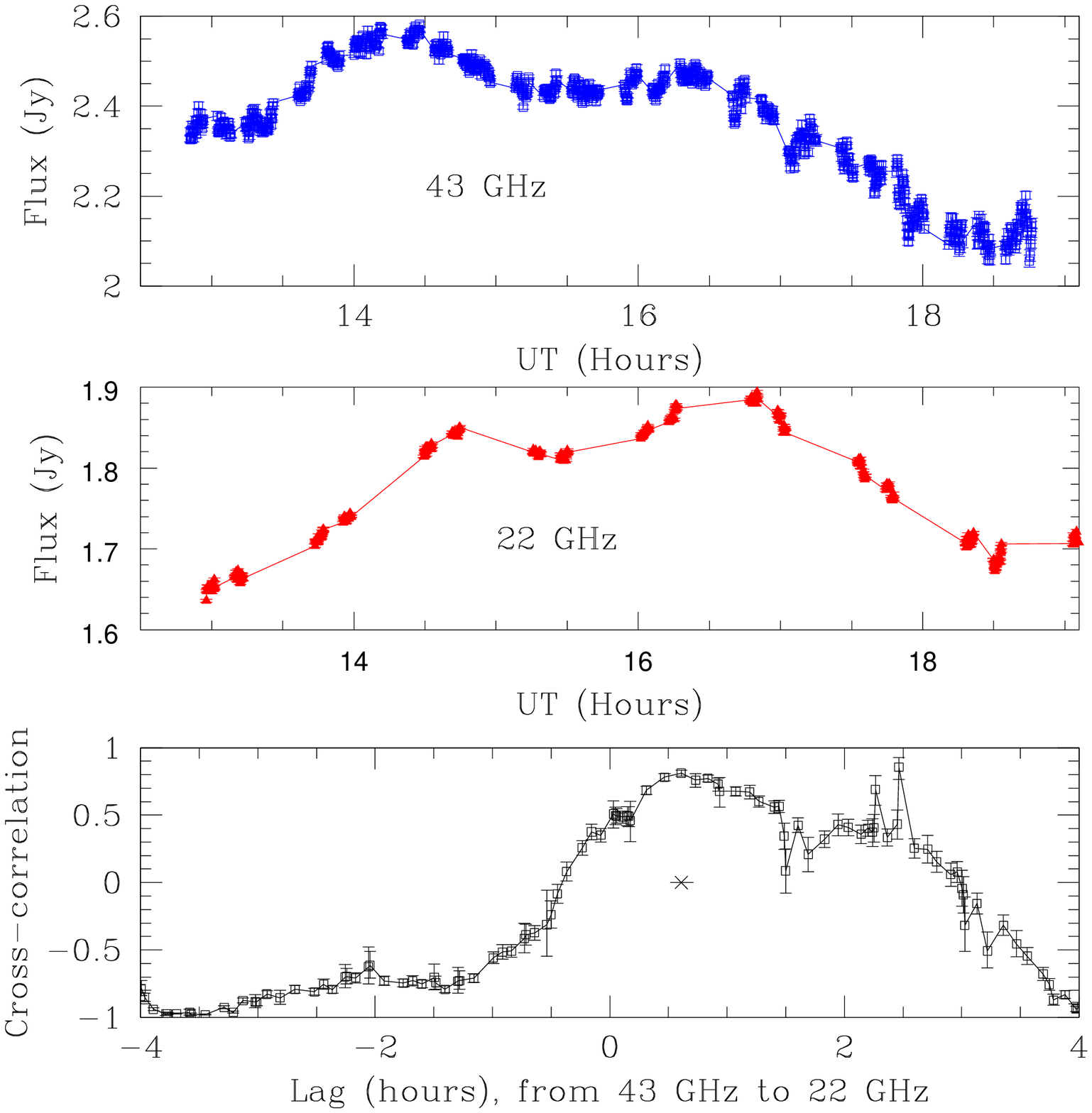}
  \includegraphics[scale=0.4,angle=0]{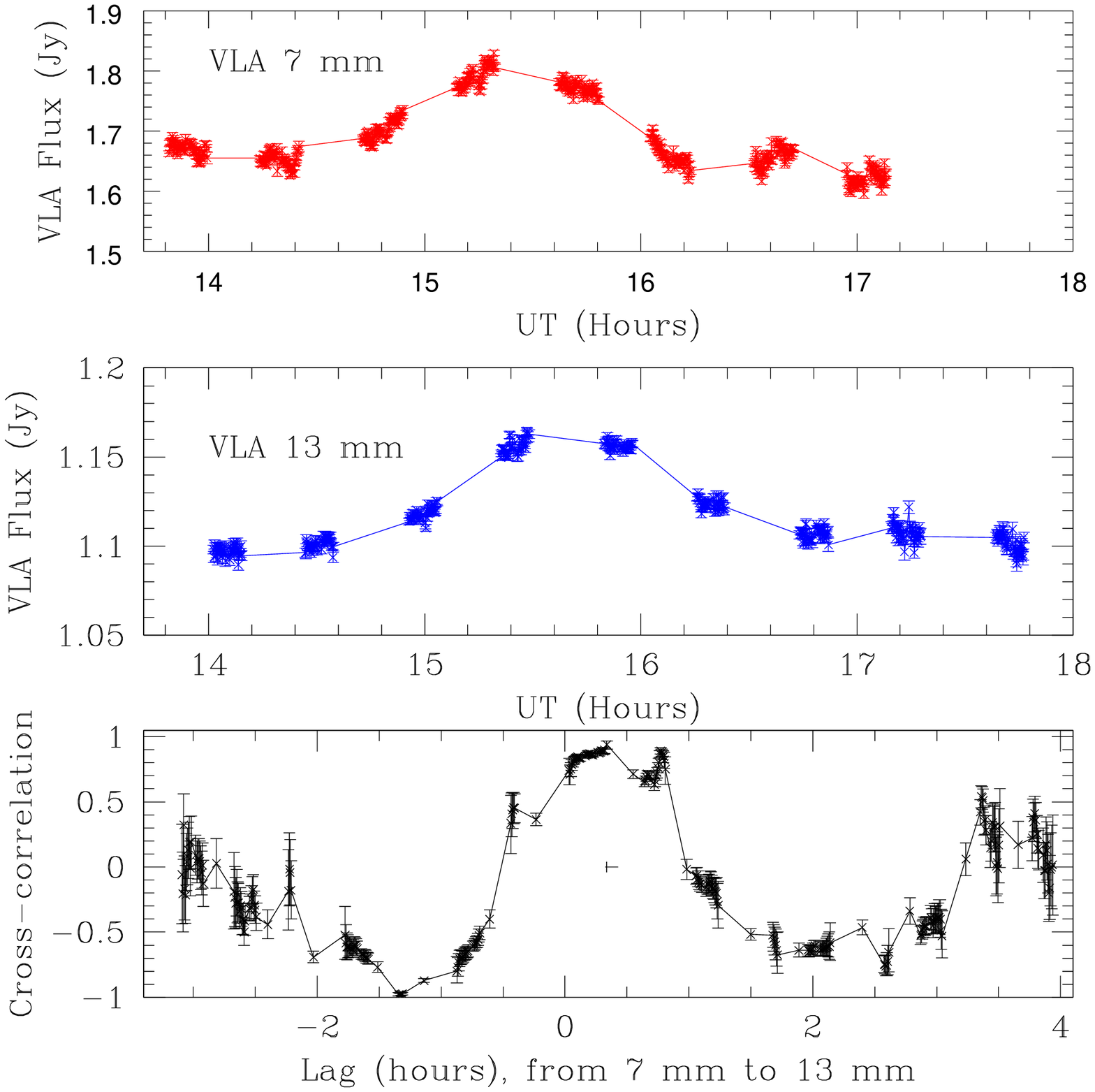}
\caption{{\bf (a) Left} Light curves of 7mm (43 GHz) and 13 mm (22 GHz) flare 
emission 
on 2006 February  10  and the corresponding cross correlation plot in the bottom 
panel. 
The 13 mm peaks show a  time  lag 
of 20.4$\pm6.8$ minutes with respect to 7 mm peaks. 
{\bf (b) Right} Similar to (a) except the data is based on 2005 February 10 data. 
The time lag is estimated to be 30$\pm12$ minutes. 
The  
maximum  likelihood peak, identified as a star with  a 1-$\sigma$ error bar,  
is  shown in each figure.  The parameters of the cross correlation peak can 
be found in Table 1. 
}
\end{figure}

\begin{figure}
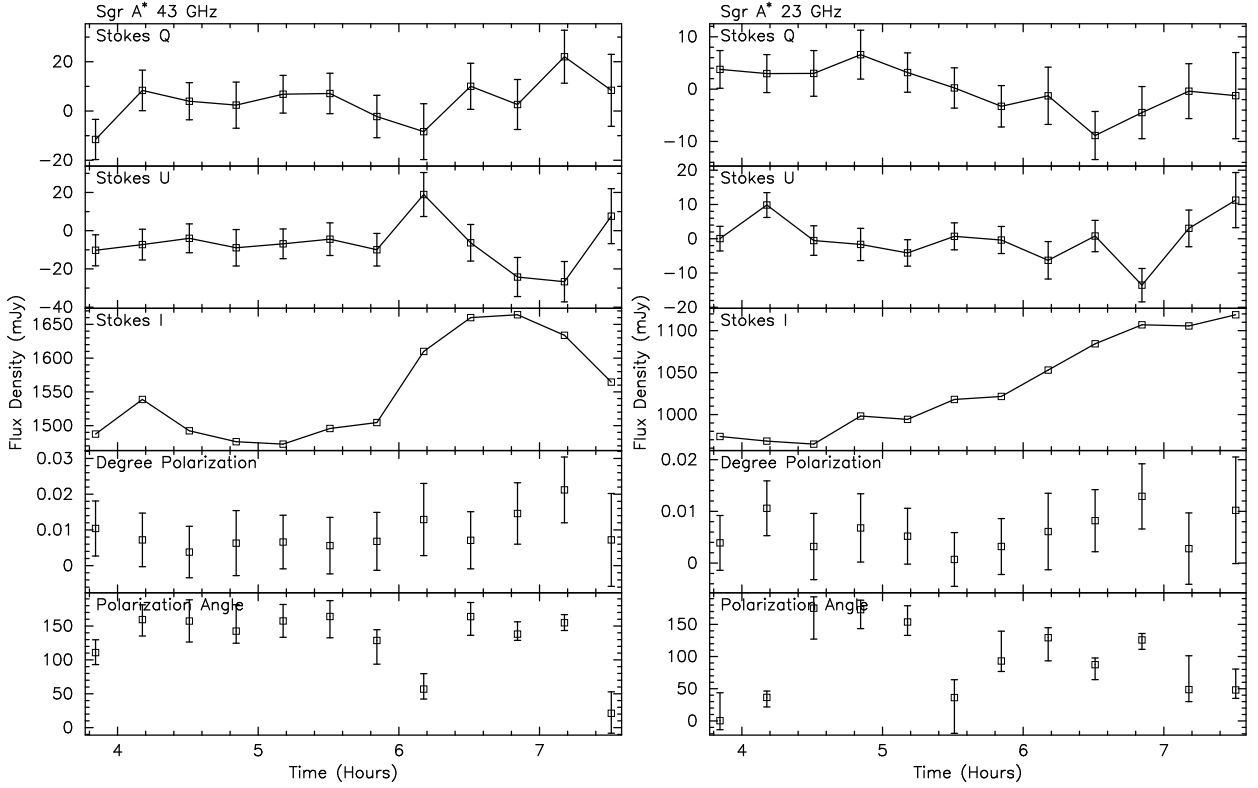

 \includegraphics[scale=0.45,angle=0]{f5a_polazn_qband.ps}
  \includegraphics[scale=0.45,angle=0]{f5b_polazn_kband.ps}
\caption{{\bf (a) Left} The light curves of Stokes Q, U, I, 
the degree of polarization and the polarization angle  at 7 mm. 
{\bf (b) Right} Similar to (a) except at 13 mm. 
The time interval sampled at 7 and 13 mm  are 1200 seconds. 
}
\end{figure}

\begin{figure}
 \includegraphics[scale=0.6,angle=0]{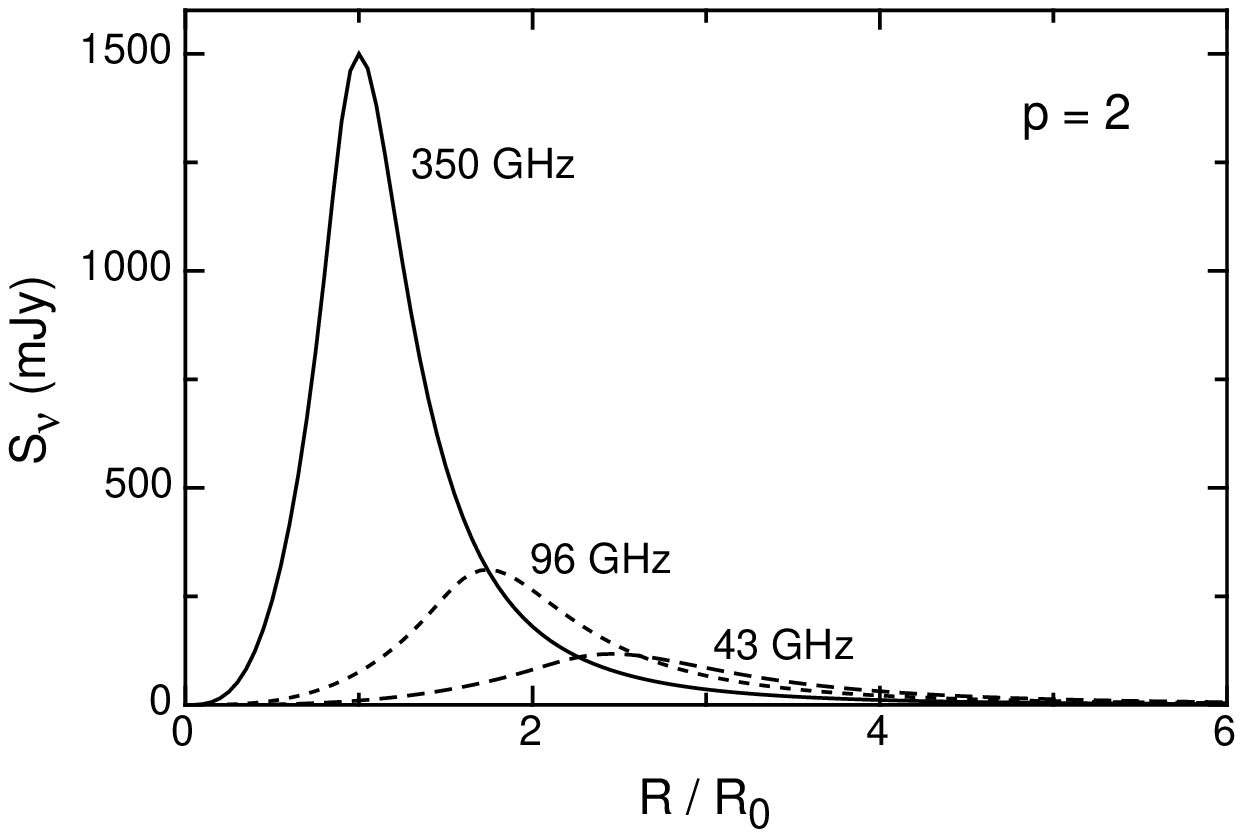}
  \includegraphics[scale=0.6,angle=0]{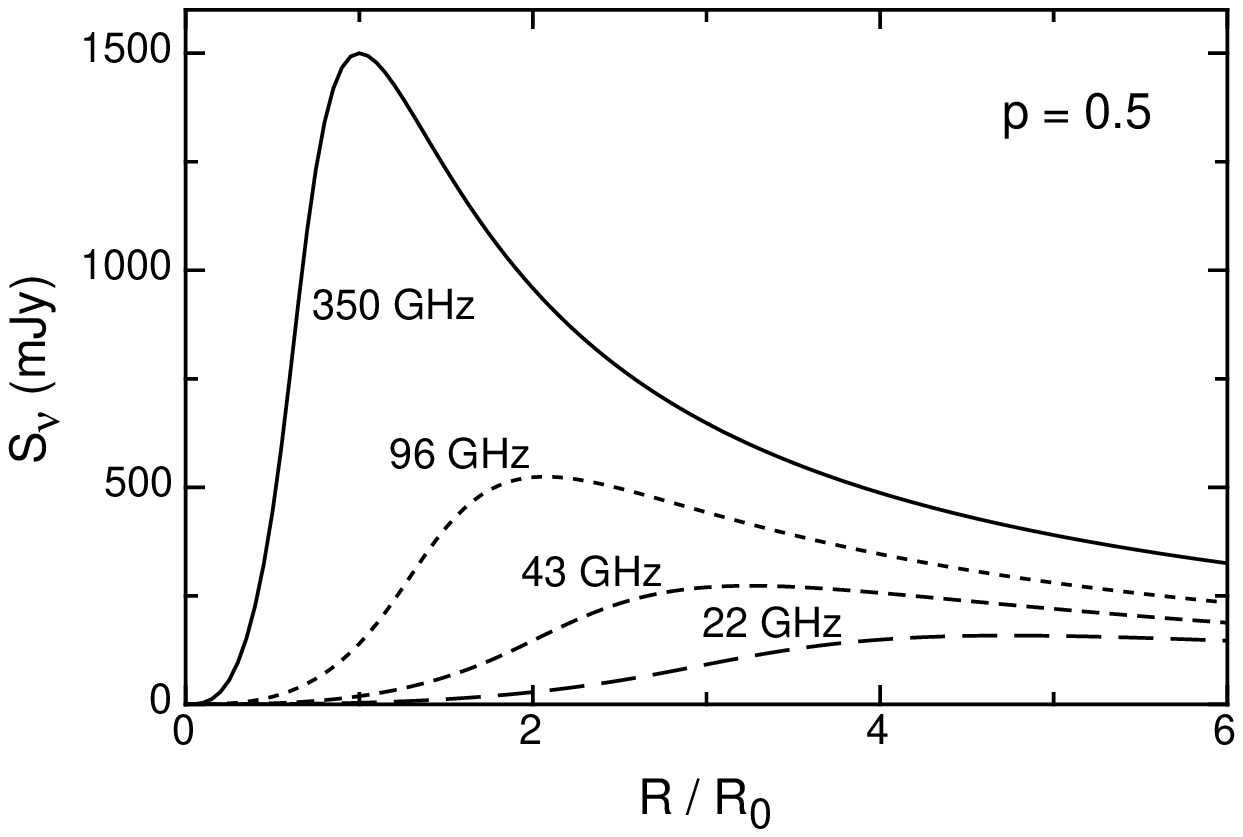}
\caption{{\bf (a) Left} Theoretical light curve of Stokes I
for optically thick synchrotron emission at submm and radio frequencies 
as a function of  expanding blob radius. These light curves assume a 
energy power law index p=2.   
{\bf (b) Right} Similar to (a) except that the particle spectrum is flat with p=0.5.  
Note the long decay time of the light curve as the spectrum flattens. 
}
\end{figure}

\begin{figure}
 \includegraphics[scale=0.6,angle=0]{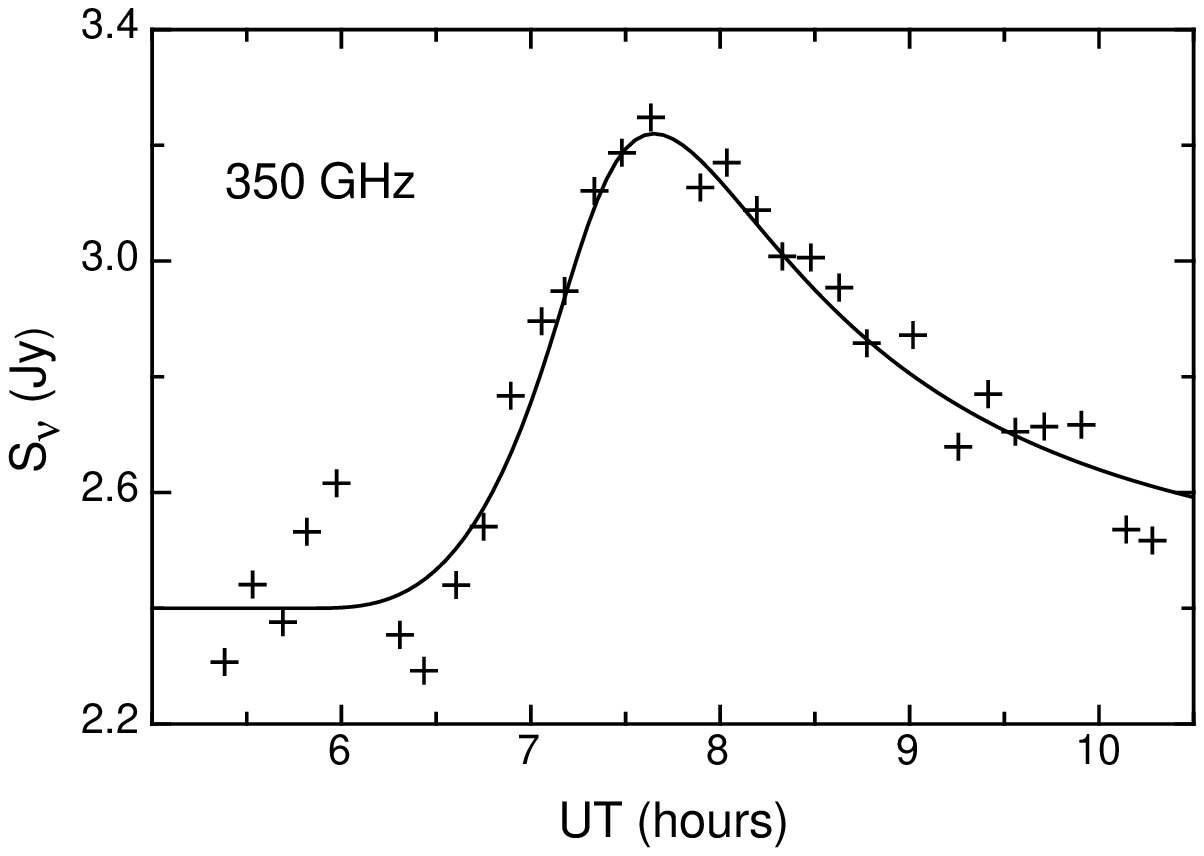}
  \includegraphics[scale=0.6,angle=0]{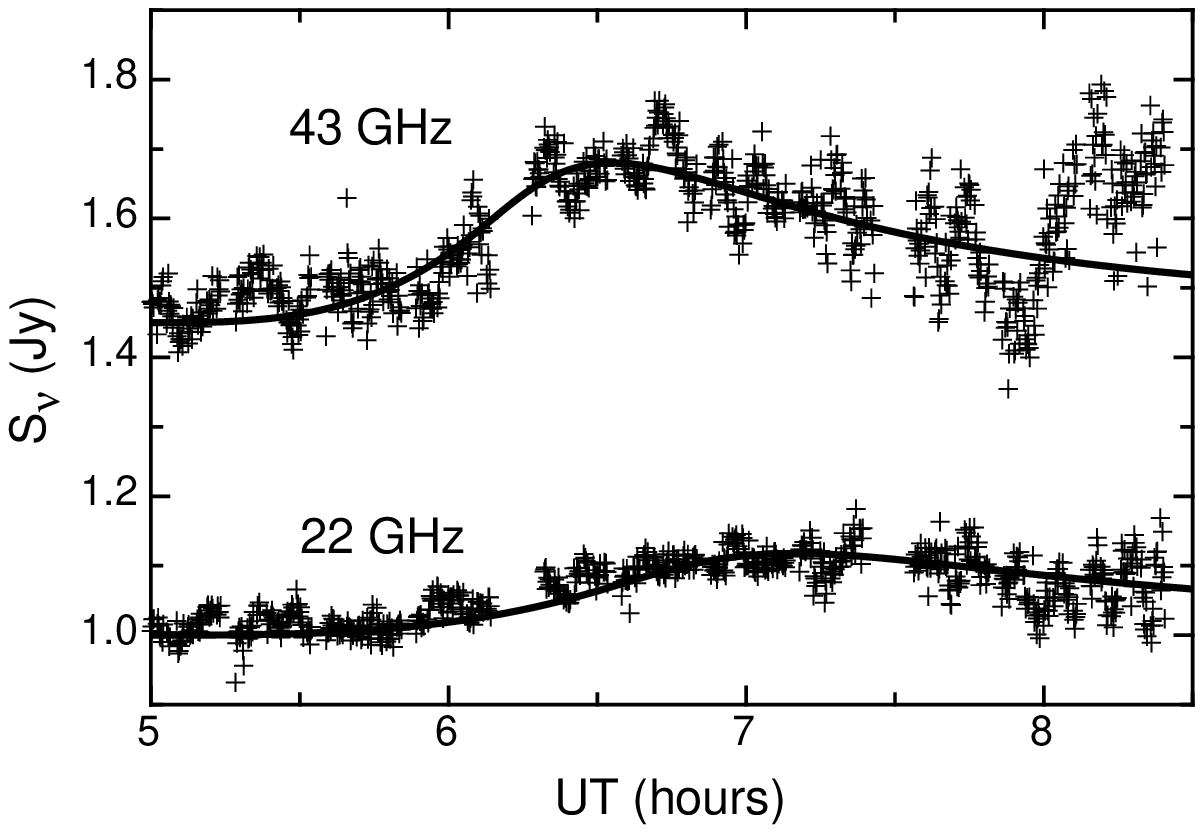}
\caption{{\bf (a) Left} The solid line represents a model fitting 
the sub-mm light curve  on 2006 July 17 at 350 GHz. Note that the fit is made only 
to the 
light curve of the bright flare and  not to the weak flare peaking around 6h UT. 
{\bf (b) Right} Similar to (a) except that
the 43 and 22 GHz 
light curves are fitted simultaneously for the 2006 July 17 
data.  
The parameters of the 2006 July 17 fits for both the sub-mm  and 
 radio light curves   can be found in Table 2. }
\end{figure}  

\begin{figure}
 \includegraphics[scale=0.6,angle=0]{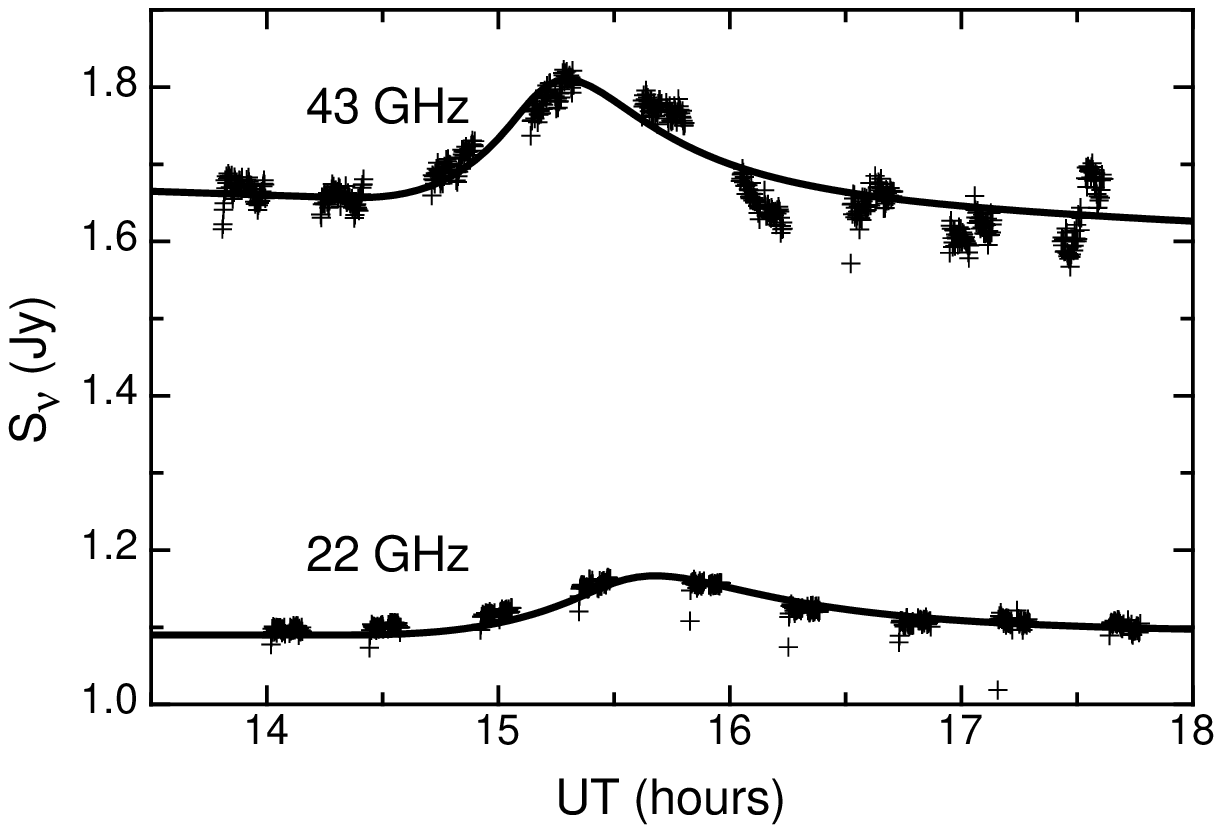}
  \includegraphics[scale=0.6,angle=0]{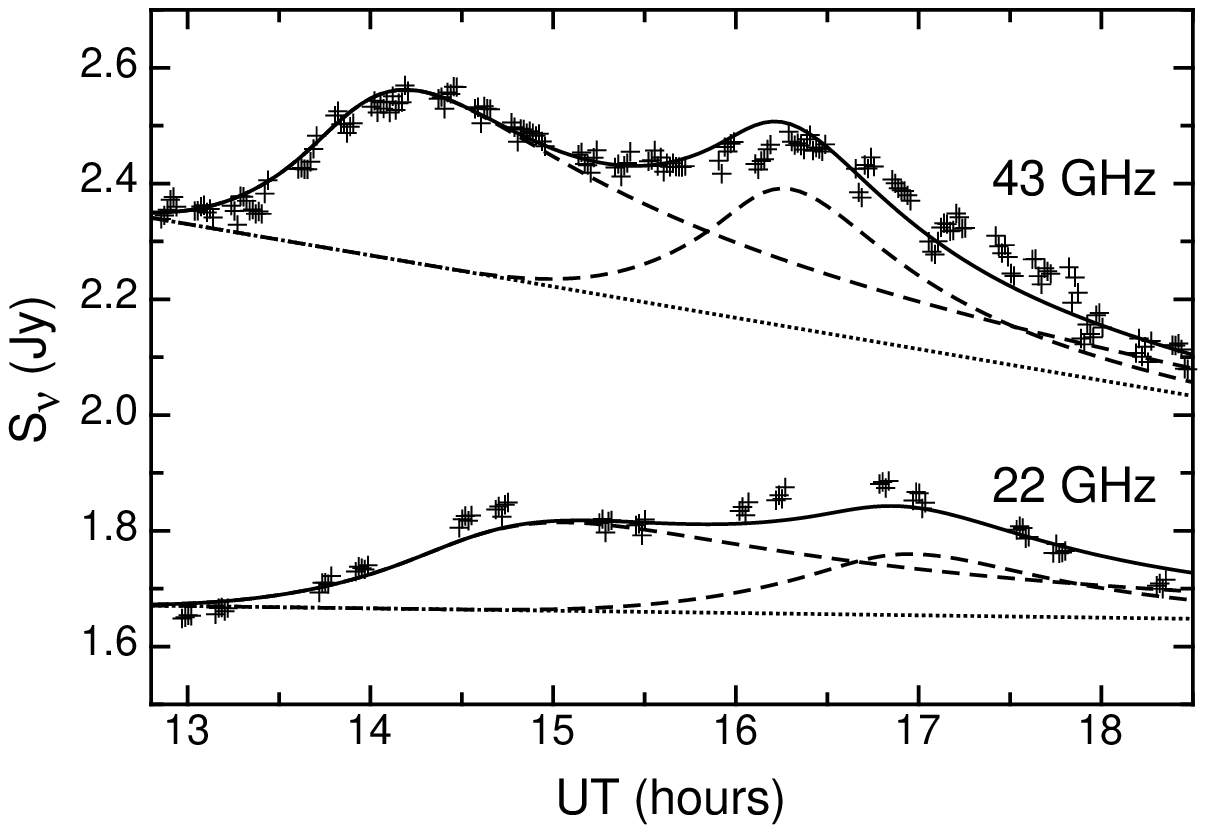}
\caption{{\bf (a) Left}  Similar to Figure 7 except that the light curve of 
 a single  flare 
at 43 and 22 GHz are fitted simultaneously. The data is taken on 
2005 February 10. 
{\bf (b) Right} Similar  to (a) except that the light curves of two  overlapping 
flares on 2006 February 
are fitted simultaneously 
at  43 and 22 GHz. 
The parameters of the fits to the 2005, February 10 and 2006 February 10  
light curves   can be found in Table 2 . }
\end{figure}  

\begin{figure}
 \includegraphics[scale=0.6,angle=0]{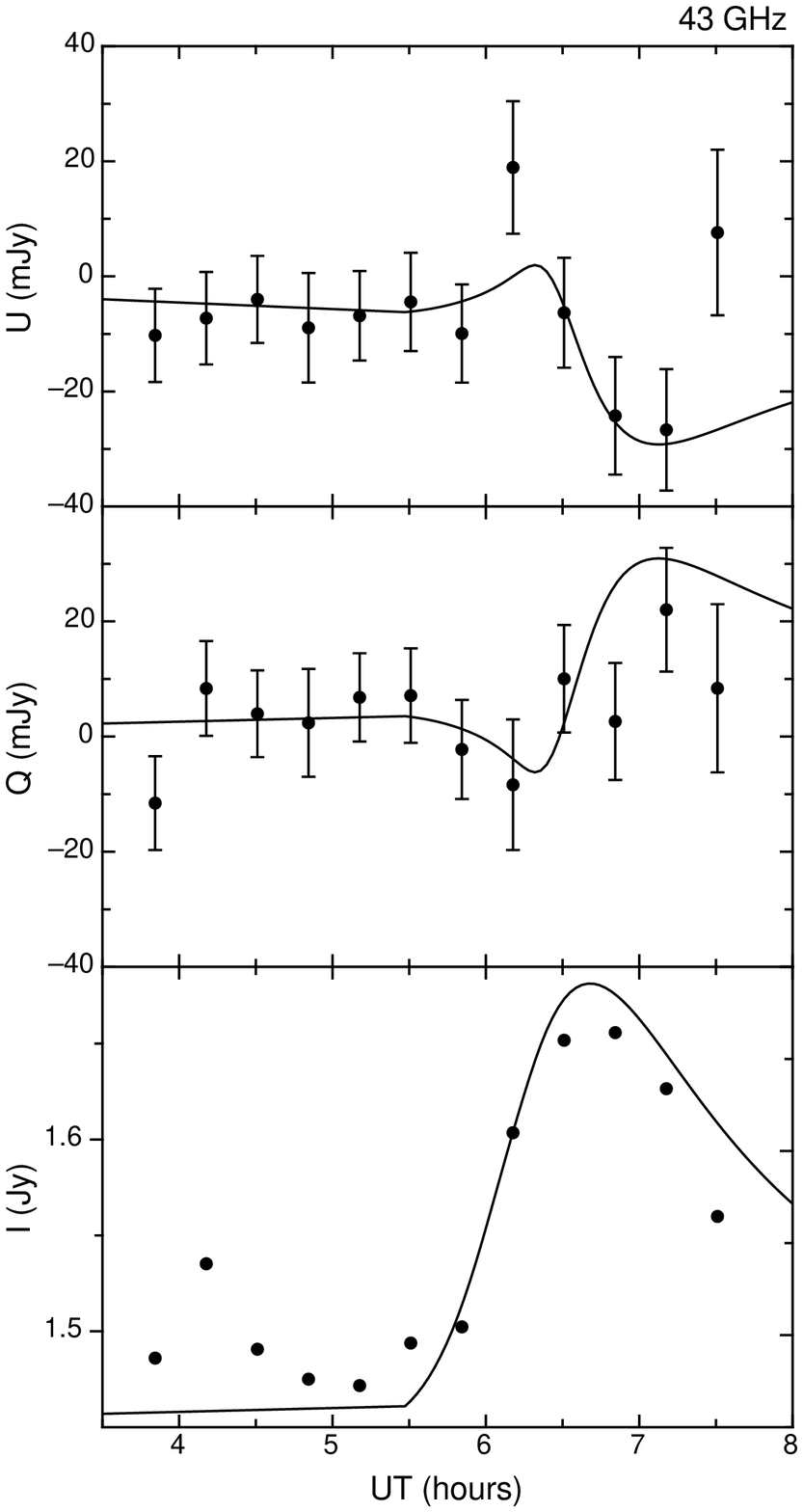}
 \includegraphics[scale=0.6,angle=0]{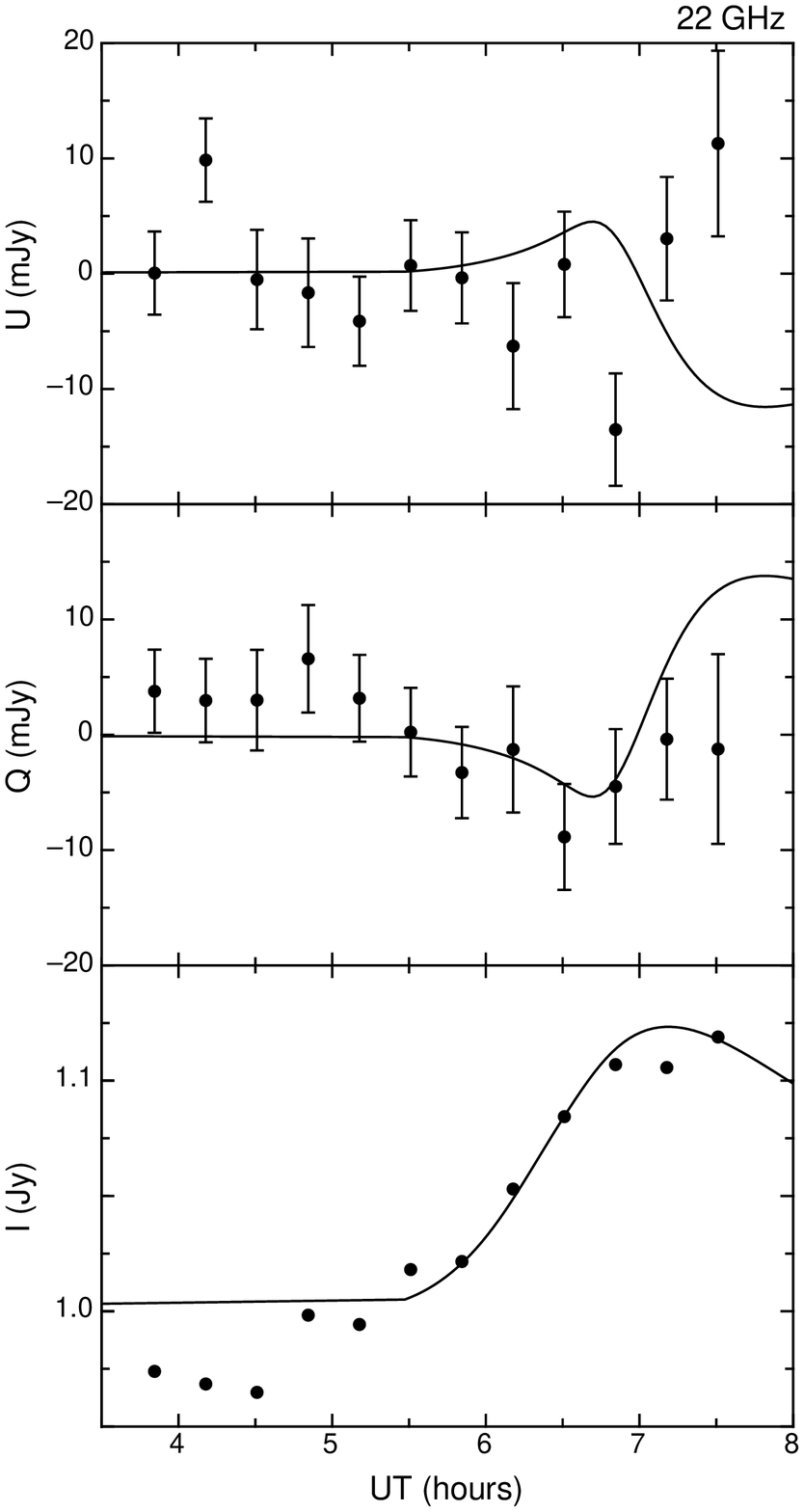}
\caption{ {\bf (a) Left} The light curves of  IQU Stokes  on 2006 July 17 at 43 
GHz 
are shown in
three panels. The solid curves show  simultaneous  fit to the light curves using the 
expanding plasmon model. {\bf (b) Right} Similar to (a) except at 22 GHz }
\end{figure}

\vfill\eject


\def\mdot   {\hbox{$\dot M$}}                   
\def\msol   {\hbox{$M_\odot$}}                  
\def\sec    {$^{\rm s}$}                        
\def\kms    {\hbox{km{\hskip0.1em}s$^{-1}$}}    
\def\kmsyr  {\hbox{km{\hskip0.1em}s$^{-1}${\hskip0.1em}yr$^{-1}$}}  
\def\i      {\hbox{\it I}}                      
\def\etal   {{\it et al. }}                     
\def\deg    {$^{\circ}$}                        
\def\amin   {$^{\prime}$}                       
\def\asec   {$^{\prime\prime}$}                 
\def\hour   {$^{\rm h}$}                        
\def\min    {$^{\rm m}$}                        
\def\sec    {$^{\rm s}$}                        
\def\ddeg   {\hbox{$.\!\!^\circ$}}              
\def\damin  {\hbox{$.\mkern-4mu^\prime$}}       
\def\dasec  {\hbox{$.\!\!^{\prime\prime}$}}     
\def\dhour  {\hbox{$.\!\!^{\rm h}$}}            
\def\dmin   {\hbox{$.\!\!^{\rm m}$}}            
\def\dsec   {\hbox{$.\!\!^{\rm s}$}}            
\def\jyperbeam {Jy{\hskip0.1em}\hbox{beam$^{-1}$}} 


\begin{deluxetable}{lccccccccc}
\small
\setlength{\tabcolsep}{0.02in} 
\tablenum{1}
\tablecolumns{9}
\tablewidth{0pt}
\tablecaption{Measured Peak Fluxes and Time Lags}
\tablehead{
\colhead{Date} &
\multicolumn{4}{c}{Peak flux} & &
\multicolumn{3}{c}{Time Delay} \\
\cline{2-5}
\cline{7-9}
\colhead{} &
\colhead{7mm} &
\colhead{13mm} &
\colhead{850$\mu$m} &
\colhead{X-ray} &
\colhead{\ } &
\colhead{X-ray--850$\mu$m} &
\colhead{7mm--13mm} &
\colhead{850$\mu$m--13mm} \\
\colhead{} &
\colhead{(Jy)} &
\colhead{(Jy)} &
\colhead{(Jy)} &
\colhead{(erg s$^{-1}$)} &
\colhead{\ } &
\colhead{(min)} &
\colhead{(min)} &
\colhead{(min)} \\
}
\startdata
2006 July 17 & 1.76 & 1.2 & 3.35 & 4$\times$10$^{34}$ & & 110 $\pm$ 17 & 20.4 $\pm$ 
6.8 & --- \\
2006 July 16 & ---  & 1.2 & 3.09 & --- & & --- & --- & 65$^{+10}_{-23}$ \\
2005 Feb 10 & 1.76 & 1.15 & --- & --- & & --- & 30 $\pm$ 12 & --- \\
2006 Feb 10  & 2.6 & 1.76 & --- & --- & & --- & 20 $\pm$ 6 & ---\\
\hfil 2006 Feb 10  \hfil  & 2.5 & 1.80 & --- & --- & & --- & 20 $\pm$ 6 & ---\\
\enddata
\end{deluxetable}



\def\mdot   {\hbox{$\dot M$}}                   
\def\msol   {\hbox{$M_\odot$}}                  
\def\sec    {$^{\rm s}$}                        
\def\kms    {\hbox{km{\hskip0.1em}s$^{-1}$}}    
\def\kmsyr  {\hbox{km{\hskip0.1em}s$^{-1}${\hskip0.1em}yr$^{-1}$}}  
\def\i      {\hbox{\it I}}                      
\def\etal   {{\it et al. }}                     
\def\deg    {$^{\circ}$}                        
\def\amin   {$^{\prime}$}                       
\def\asec   {$^{\prime\prime}$}                 
\def\hour   {$^{\rm h}$}                        
\def\min    {$^{\rm m}$}                        
\def\sec    {$^{\rm s}$}                        
\def\ddeg   {\hbox{$.\!\!^\circ$}}              
\def\damin  {\hbox{$.\mkern-4mu^\prime$}}       
\def\dasec  {\hbox{$.\!\!^{\prime\prime}$}}     
\def\dhour  {\hbox{$.\!\!^{\rm h}$}}            
\def\dmin   {\hbox{$.\!\!^{\rm m}$}}            
\def\dsec   {\hbox{$.\!\!^{\rm s}$}}            
\def\jyperbeam {Jy{\hskip0.1em}\hbox{beam$^{-1}$}} 


\begin{deluxetable}{lcccccccc}
\small
\setlength{\tabcolsep}{0.04in} 
\tablenum{2}
\tablecolumns{9}
\tablewidth{0pt}
\tablecaption{Fitted Parameters to the Light Curves}
\tablehead{
\colhead{Date} &
\colhead{Quiescent} &
\multicolumn{3}{c}{Background subtracted peak flux} &
\colhead{Particle} &
\colhead{Expansion} &
\colhead{Initial} &
\colhead{Initial} \\
\cline{3-5}
\colhead{} &
\colhead{Flux} &
\colhead{7mm} &
\colhead{13mm} &
\colhead{850$\mu$m} &
\colhead{Index} &
\colhead{Speed} &
\colhead{B} &
\colhead{Radius} \\
\colhead{} &
\colhead{Jy} &
\colhead{(mJy)} &
\colhead{(mJy)} &
\colhead{(mJy)} &
\colhead{p} &
\colhead{v/c} &
\colhead{(Gauss)} &
\colhead{(R$_s$)}\\
}
\startdata
2005 Feb 10 & 1.62--1.64 & 160 & --- & --- & 1.5 & 0.018 & 12 & 1.9\\
\hfil $''$ \hfil & 1.09 & --- & 75 & --- & 1.5 & 0.018 & 12 & 1.9 \\

2006 Feb 10 &2.06& 300 & --- & --- & 1 & 0.013 & 10.7 & 2.5  \\
\hfil $''$ \hfil & 2.33& 240 & --- & --- & 2 & 0.016 & 10.4 & 3.2  \\
\hfil $''$ \hfil &1.65& --- & 150 & --- & 1 & 0.013 & 10.7 & 2.5  \\
\hfil $''$ \hfil & 1.67& --- & 110 & --- & 2 & 0.016 & 10.4 & 3.2  \\

2006 July 17 & 1.45 & 230 & --- & --- & 1 & 0.15  & 11   & 2.2 \\
\hfil $''$ \hfil &1.0 & --- &  120 & ---  &  $''$ & $''$ &  $''$ & $''$  \\
\hfil $''$ \hfil &2.40 & --- & --- & 820 & 1 & 0.0028 & 76 & 0.50  \\
\enddata
\end{deluxetable}


\begin{thebibliography}{99}
\bibitem[1]{a} Alexander, T.  1997 in Astronomical Time Series, ed. D. Maoz, A. Sternberg \& E. 
Leibowitz  (Dordrecht:Kluwer), 163

\refitem Baganoff, F. K., Bautz, M. W., Brandt, W. N., Chartas, G., Feigelson, E. D., Garmire, G. P., Maeda, Y., Morris, M., Ricker, G. R., Townsley, L. K. \& Walter, F. 2001, 
         Nature, 413, 45
		 
\bibitem[1]{a} 
Baganoff, F. K.,  Maeda, Y.,  Morris, M.,  Bautz, M. W.,  Brandt, W. N.,  Cui, W., 
et al.  2003, ApJ, 591, 891	

\bibitem[4]{a} Ballantyne, D. R., Ozel, F. \& Psaltis, D. 2007,
 ApJ, 663, L17

 
\refitem B\'elanger, G., Goldwurm, A., Melia, F., Ferrando, P., Grosso, N., Porquet, D., Warwick, R. \& Yusef-Zadeh, F. 2005, 
         ApJ, 635, 1095
		 
\refitem Bittner, J. M., Liu, S., Fryer, C. L. \& Petrosian, V. 2007, 
 ApJ, 661, 863
 
\refitem Blandford, R. D. \& Begelman, M. C. 1999, 
         MNRAS, 303, L1
		 
\bibitem[4]{a} 	Blandford, R. \&  Eichler, D. 1987, PhR, 154, 1	

\bibitem[4]{a} Bower, G. C. Falcke, H., Wright, M. C. \& Backer, D. C. 
 2005, \apjl, 618, L29

\refitem Bower, G. C., Wright, M. C. H., Falcke, H. \& Backer, D. C. 2003, 
         ApJ, 588, 331
		 

\bibitem[2]{a} 
Drury, L. O. C. \& V\"olk, J. H. 1981,  \apj, 248, 344


\refitem Eckart, A., Baganoff, F. K., Morris, M., Bautz, M. W., Brandt, W. N., Garmire, G. P., Genzel, R., Ott, T., Ricker, G. R., Straubmeier, C., Viehmann, T., Sch\"odel, R., Bower, G. C. \& Goldston, J. E. 2004, 
         A\&A, 427, 1
\refitem Eckart, A., Baganoff, F. K., Schoedel, R., Morris, M., Genzel, R., Bower, G. C., Marrone, D., Moran, J. M., Viehmann, T., Bautz, M. W., Brandt, W. N., Garmire, G. P., Ott, T., Trippe, S., Ricker, G. R., Straubmeier, C., Roberts, D. A., Yusef-Zadeh, F., Zhao, J. H. \& Rao, R. 2006, 
         A\&A, 450, 535
 
\bibitem[7]{a} Edelson, R. A. \& Krolik, J. H. 1988, ApJ, 333, 646	

\refitem Eisenhauer, F., Genzel, R., Alexander, T., Abuter, R., Paumard, T., Ott, T., 
         Gilbert, A., Gillessen, S., Horrobin, M., Trippe, S., Bonnet, H., Dumas, C., 
		 Hubin, N., Kaufer, A., Kissler-Patig, M., Monnet, G., Str\"obele, S., Szeifert, T., 
		 Eckart, A., Sch\"odel, R. \& Zucker, S. 2005, 
         ApJ, 628, 246
		 
\bibitem[7]{a} Falcke, H., Goss, W. M.,  Matsuo, H.,  Teuben, P.,  Zhao, 
Jun-Hui \& Zylka, R. 1998, ApJ, 499,731	

\bibitem[7]{a} 
Falcke, H. \& Markoff, S.  2000, \aap, 362, 113

\bibitem[7]{a} 
Garmire, G. P.,  Bautz, M. W.,  Ford, P. G.,  Nousek, J. A.,  Ricker, G. R., Jr. 
2003, SPIE, 4851, 28G	



\bibitem[10]{a} 
Ghez, A. M.,  Wright, S. A.,  Matthews, K., Thompson, D.,
Le Mignant, D.,  Tanner, A., Hornstein, S. D.,  Morris, M.,
Becklin, E. E. \& Soifer, B. T. 2004, ApJ, 601, L159

\refitem Gillessen, S., Eisenhauer, F., Quataert, E., Genzel, R., Paumard, T., Trippe, S., Ott, T., Abuter, R., Eckart, A., Lagage, P. O., Lehnert, M. D., Tacconi, L. J. \& Martins, F. 2006, 
 ApJ, 640, L163
 
\refitem Goldston, J. E., Quataert, E. \& Igumenshchev, I. V. 2005, 
 ApJ, 621, 785

\refitem Igumenshchev, I. V., Narayan, R. \& Abramowicz, M. A. 2003, 
         ApJ, 592, 1042
		 
\bibitem[10]{a} 
Jones, T. W. \&  O'Dell, S.L. 1977, ApJ, 214, 522
	
\bibitem[10]{a}  Hornstein, S. D., Matthews, K.,  Ghez, A. M.,  Lu, J. R.,  Morris, M.,  Becklin, E. 
E.,  Rafelski, M. \&   Baganoff, F. K. 2007, ApJ, 667, 900

\bibitem[10]{a} 
Liu, S. \&  Melia, F. 2001, ApJ, 561, L77

\bibitem[10]{a} 
Liu, S. \&  Melia, F. 2002, ApJ, 573, L23

\refitem Liu, S., Petrosian, V. \& Melia, F. 2004, 
 ApJ, 611, L101

\refitem Liu, S., Melia, F. \& Petrosian, V. 2006a, 
 ApJ, 636, 798

\refitem Liu, S., Petrosian, V., Melia, F. \& Fryer, C. L. 2006b, 
 ApJ, 648, 1020


\bibitem[13]{a} Macquart, J. P.  \& Bower, G. C.  2006, \apj, 641, 302


\bibitem[14]{a} Marrone, D. P., Moran, J. M., Zhao, J.-H. \& Rao, R. 2006, \apj, 640, 
308

\refitem Marrone, D. P., Baganoff, F. K., Morris, M., Moran, J. M., Ghez, A. M., Hornstein, S. D., 
         Dowell, C. D., Munoz, D. J., Bautz, M. W., Ricker, G. R., Brandt, W. N., Garmire, G. P., 
		 Lu, J. R., Matthews, K., Zhao, J.-H., Rao, R. \& Bower, G. C. 2007, 
         ApJ, submitted
		 
\bibitem[15]{a} Mauerhan, J. C., Morris, M., Walter, F.  \& Baganoff, F.K.  2005, \apjl,
623, L25
 
\refitem Melia, F. 1992 
         ApJ, 387, L25
		 
\bibitem[16]{a} Melia, F.  \& Falcke, H.  2001, \araa, 39, 309

\refitem Melrose, D. B. \& Pope, M. H. 1993, 
         PASA, 10, 222

\bibitem[17]{a} Miyazaki, A., Tsutsumi, T. \& Tsuboi, M.  2004, \apjl, 611, L97

\refitem Narayan, R., Mahadevan, R., Grindlay, J. E., Popham, R. G. \& Gammie, C. 1998, 
         ApJ, 492, 554
		 
\refitem Pope, M. H. \& Melrose, D. B. 1994, 
         PASA, 11, 175
\bibitem[19]{a} Reid, M. J.  \& Brunthaler, A.  2004, \apj, 616, 872

\bibitem[20]{a} Sch{\"o}del, R.,  Ott, T., Genzel, R., Eckart, R. D., Mouawad, N. \&
Alexander, T.  2003, \apj, 596, 1015

\bibitem[25]{a} Serabyn, E., Carlstrom, J.,  Lay, O., Lis, D. C., Hunter, T. R., 
Lacy,  J.H. \& Hills, R. E. 1997, ApJ, 490, L77

\refitem Sharma, P., Quataert, E. \& Stone, J. M. 2007, 
 ApJ, in press
 
\refitem Shen, Z.-Q., Lo, K. Y., Liang, M.-C., Ho, P. T. P. \& Zhao, J.-H. 2005, 
         Nature, 438, 62

\bibitem[22]{a} van der Laan, H.  1966, \nat, 211, 1131

\refitem Weisskopf, M. C., Brinkman, B., Canizares, C., Garmire, G., Murray, S. \& Van Speybroeck, L. P. 2002, 
         PASP, 114, 1
		 
\refitem Yuan, F., Markoff, S. \& Falcke, H. 2002, 
         A\&A, 383, 854

\refitem Yuan, F., Quataert, E. \& Narayan, R. 2003, 
		 ApJ, 598, 301
 
\refitem Yuan, F., Quataert, E. \& Narayan, R. 2004, 
         ApJ, 606, 894
		 
\bibitem[23]{a} Yusef-Zadeh, F.,  Bushouse, H., Dowell, C. D.,  Wardle, M., Roberts, D.,  Heinke, 
C.,  Bower, G. C.,  Vila-Vilar, B.,  Shapiro, S.,  Goldwurm, A., 
 Belanger, G.,  2006a, \apj, 644, 198
 
\bibitem[24]{a} Yusef-Zadeh, F., Morris, M. \& Ekers, R. D.  1990, \nat, 348, 45

\bibitem[25]{a} 
Yusef-Zadeh, F. Roberts, D., Wardle, M., Heinke, C. O. \&
Bower, G. C.   2006b, \apj, 650, 189

\bibitem[25]{a} Yusef-Zadeh, F.,  Wardle, M.,  Cotton, W. D.,  Heinke, C. O. \& Roberts, D. A.
2007, ApJ, 668, L47



\bibitem[25]{a} Zylka, R., Mezger, P.G., Ward-Thompson, D., Duschl, W. J. \& 
Lesch, H. 1995, A\&A, 297, 83

\bibitem[26]{a} Zhao, J.-H., Herrnstein, R. M., Bower, G. C., Goss, W. M. \&  
Liu, S.M.  2004, \apjl, 603, L85

\bibitem[]{}

\end{thebibliography}
\end{document}